\documentclass[pre,twocolumn,showpacs]{revtex4-1}
\usepackage{amsmath}
\usepackage{bbm}
\usepackage{graphicx}
\DeclareGraphicsExtensions{.pdf,.jpg,.eps}
\usepackage{color}

\begin{document}

\title{Differential growth of wrinkled biofilms} 

\author{D. R. Espeso}
\affiliation{\small Centro Nacional de Biotecnolog\'{\i}a, CSIC, Madrid 28049, Spain}

\author{A. Carpio}
\email{Corresponding author: carpio@mat.ucm.es}
\affiliation{\small Departamento de Matematica Aplicada, Universidad Complutense 28040, Madrid, Spain}

\author{B. Einarsson}
\affiliation{\small Center for Complex and Nonlinear Science, UC Santa Barbara, California 93106, USA}

\date{Feb 2015, published in Physical Review E 91, 022710 (2015)}

\begin{abstract}
Biofilms are antibiotic-resistant bacterial aggregates that grow 
on moist surfaces and {can} trigger hospital-acquired infections.
They provide a classical example in biology where the 
dynamics of cellular communities may be observed and studied.
{Gene expression} regulates cell division and differentiation,
{which affect the biofilm architecture}.
Mechanical and chemical processes shape the resulting 
structure. We gain insight into the interplay between cellular 
and mechanical processes during biofilm development {on 
air-agar interfaces} by means of a hybrid 
model. Cellular behavior is governed by stochastic rules informed 
by a cascade of concentration fields for nutrients, waste and 
{autoinducers}. Cellular differentiation and death alter the 
{structure} and the mechanical properties of the biofilm, 
which is deformed according to F\"oppl-Von K\'arm\'an equations 
informed by cellular processes and the interaction with the substratum. 
Stiffness gradients due to growth {and swelling} 
produce wrinkle branching. We are able to reproduce wrinkled 
structures often formed by biofilms on air-agar interfaces, as well as 
spatial distributions of differentiated cells commonly observed 
with {\em B. subtilis}. 
\end{abstract}


\maketitle

\section{Introduction}

From bacterial communities to multicellular tissues, three dimensional biological structures emerge through {poorly understood} interactions between mechanical forces and cellular processes. {Being apparently less complex than multicellular organisms, bacterial biofilms may provide model systems for analyzing the interplay between cellular and mechanical features of three dimensional self-organization during growth.}
 
Biofilms are bacterial aggregates that form on moist surfaces and are sheltered from external aggressions by a self-produced extracellular matrix (ECM) made of exopolymeric substances (EPS) \cite{development,matrix2}. This matrix makes them uncommonly resistant to antibiotics, disinfectants, flows and other chemical or mechanical agents \cite{hoiby}. Biofilms are involved in chronic infections associated to biomedical implants \cite{costerton,darouiche} and in many industrial problems, such as biocorrosion, biofouling, efficiency reduction in heat exchange systems and food poisoning  \cite{carvalho,eguia,xiong}. They are beneficial in bioremediation and biocontrol \cite{singh,slimy} or in wastewater treatments \cite{wuertz}.

Biofilm structure depends among other variables on the bacterial strain and the nutrient source, but is also influenced by environmental conditions. Biofilms grown in flows \cite{purevdorj,stonepnas} differ noticeably from biofilms expanding on air-solid or air-liquid interfaces not exposed to a shear force \cite{hera,osmotic}. These vary depending on the surface, as it is the case for {\em Bacillus subtilis} forming pellicles on air-fluid interfaces \cite{trejo} or wrinkled films on agar \cite{asally}. 
Pattern formation is crucial to the development of biological systems.
In cellular organisms, growth is a complex process implicating biochemical and physical mechanisms occurring at a variety of length and timescales \cite{coen}. On one side, genetic programs govern cellular processes, such as growth and differentiation. On the other, mechanical properties and physical forces induce macroscopic movements of cell populations and chemical compounds \cite{asally,mammoto}. Detailed models trying to describe biofilm evolution on surfaces should take into account the interaction of these processes. 

{Many models have been proposed for biofilm expansion in flows \cite{bookmodels, klappersiam}, focussing typically on growth due to nutrient consumption. 
The standard geometry consists of a biofilm attached to a solid substratum submerged in a fluid. Growth is limited by diffusion of oxygen and nutrients from the surrounding flow into the biofilm. EPS production favors vertical spread, pushing cells into the flow to improve access to nutrients and oxygen \cite{development,xavierupwards}. Depending on the nutrient and oxygen availability, the bacterial strain and the hydrodynamic conditions,  biofilms evolve to form a variety of patterns: mushrooms, streamers, mounds, ripples and patchy shapes are observed \cite{purevdorj,xavierspread,stonepnas,pre}.
The volume fraction of EPS in these biofilms is large \cite{stonepnas}. Some models view them as cells embedded in a fluid EPS matrix \cite{picioreanueps}.  Experimental observations and measurements suggest that biofilms behave as viscoelastic materials, i.e.,  exhibit  elastic solid-like response to short time scale stimuli and viscous fluid-like response to long time scale stimuli \cite{klapperviscoelastic}. 
Two-phase flow models of biofilm spread in fluids \cite{cogantwophase} coexist with models that consider the growing biofilm an hyperelastic material \cite{dupinhyperelastic}. The interplay of biofilm filaments with the flow, however, is described as interaction between an elastic structure and a fluid \cite{picioreanustreamer,stoneelastic}. These models represent the microbial community
as a continuum. To investigate the influence of interactions at bacterium level, `agent' based models have been developed. Bacteria are described as individuals whose behavior
responds to external continuum fields. Different modeling frameworks characterize individual bacteria in different ways. Cellular automata \cite{hermanovic,laspidoucellular,kapelloshierarchical}, cellular Potts \cite{poplawski} and individual-based \cite{picioreanueps,xavierupwards,picioreanusurvey} representations  have been proposed for biofilms in flows. We will comment on the general approaches  in Section II. Basic mechanisms incorporated in them include biofilm expansion due to cell division, phenomenological descriptions of EPS production and biofilm erosion caused by the external flow. }

{Biofilms grown in other environments show a different structure and undergo different processes. Understanding their evolution requires different models, that must take into account the specific behavior of the selected bacterial strain.}  
We consider here the development of {\em B. subtilis} biofilms on air-agar interfaces, for which detailed experimental  evidence is becoming available \cite{hera,osmotic,asally,seminara,wilking}. {Recent observations reveal  the role of cellular behavior and inner mechanical processes on the biofilm constitution. These biofilms extract nutrients and water from the agar substratum itself. Cells are densely packed, glued together by a small volume fraction of extracellular substances \cite{seminara}.   The specific autoinducer sequence triggering EPS secretion has been identified \cite{hera}.  
EPS production favors water absorption and horizontal spread \cite{seminara}.  
Localized cell death leads to mechanical stress relief through buckling \cite{asally}. Instead of networks of vertical mushrooms or filaments undulating with the external flow, wrinkled films are observed \cite{osmotic,asally,wilking}.  Partial aspects of this picture have been modeled. Ref. \cite{seminara} analyzes accelerated horizontal expansion due to EPS production and swelling by means of thin film approximations calibrated to experiments.
A reaction-diffusion model for the density of living cells combined with equations for the accumulation of EPS and waste provides understanding of the cell death patterns  
in Ref. \cite{asally}.}
We introduce here a hybrid framework that might be suitable to simulate the interplay between the mechanical and cellular processes identified so far. We couple a stochastic treatment of individual cellular activities with a continuum description of elastic deformations. To facilitate the coupling of stochastic cell division, death, differentiation and swelling processes with macroscopic elastic deformations and the description of the complex interaction with the substratum, we have represented bacteria as cellular automata. Cells respond to a number of concentration fields (nutrients, waste, secreted substances), and are affected by water migration processes and elastic deformations resulting of inner stresses due to growth, swelling and the interaction with the substratum. Our numerical simulations suggest that the wrinkle branching observed in experiments may be the result of elastic deformations triggered by stiffness gradients.

\begin{figure}[!ht]
\centering
\includegraphics[width=8.5cm]{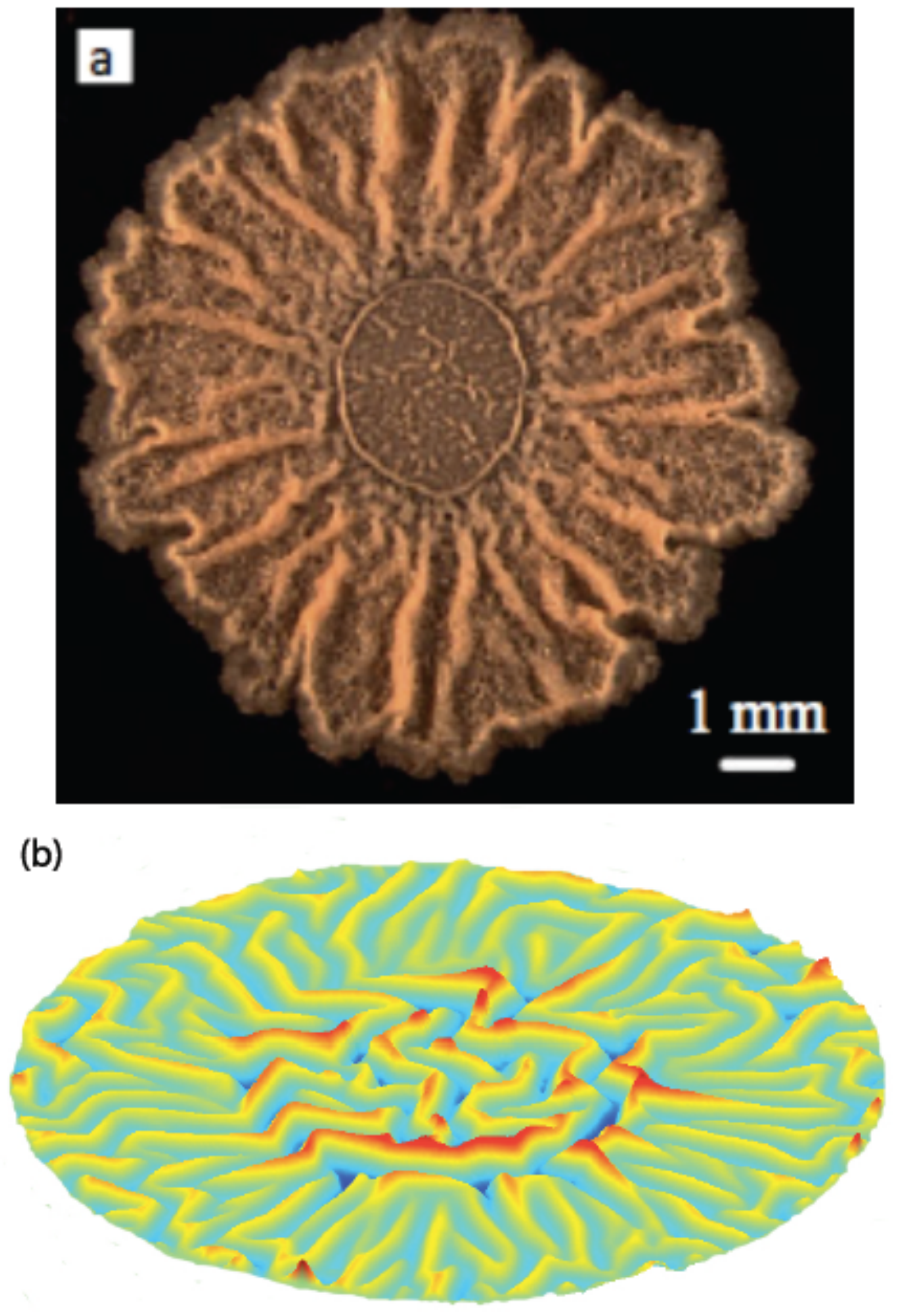}   
\caption{(Color online) 
(a) {\it B. subtilis} biofilm grown on a 1.5\% agar surface in a biofilm-inducing medium. Photograph taken after four days. Reprinted with the permission of Cambridge University Press from 
Chai L, Vlamakis H, Kolter R, MRS Bulletin,  36: 374-379 (2011),
\cite{hera}. (b) In silico film showing a corona of radial wrinkles emerging from a central core.}
\label{fig1}
\end{figure}

\begin{figure}[!ht]
\centering
\includegraphics[width=8.5cm]{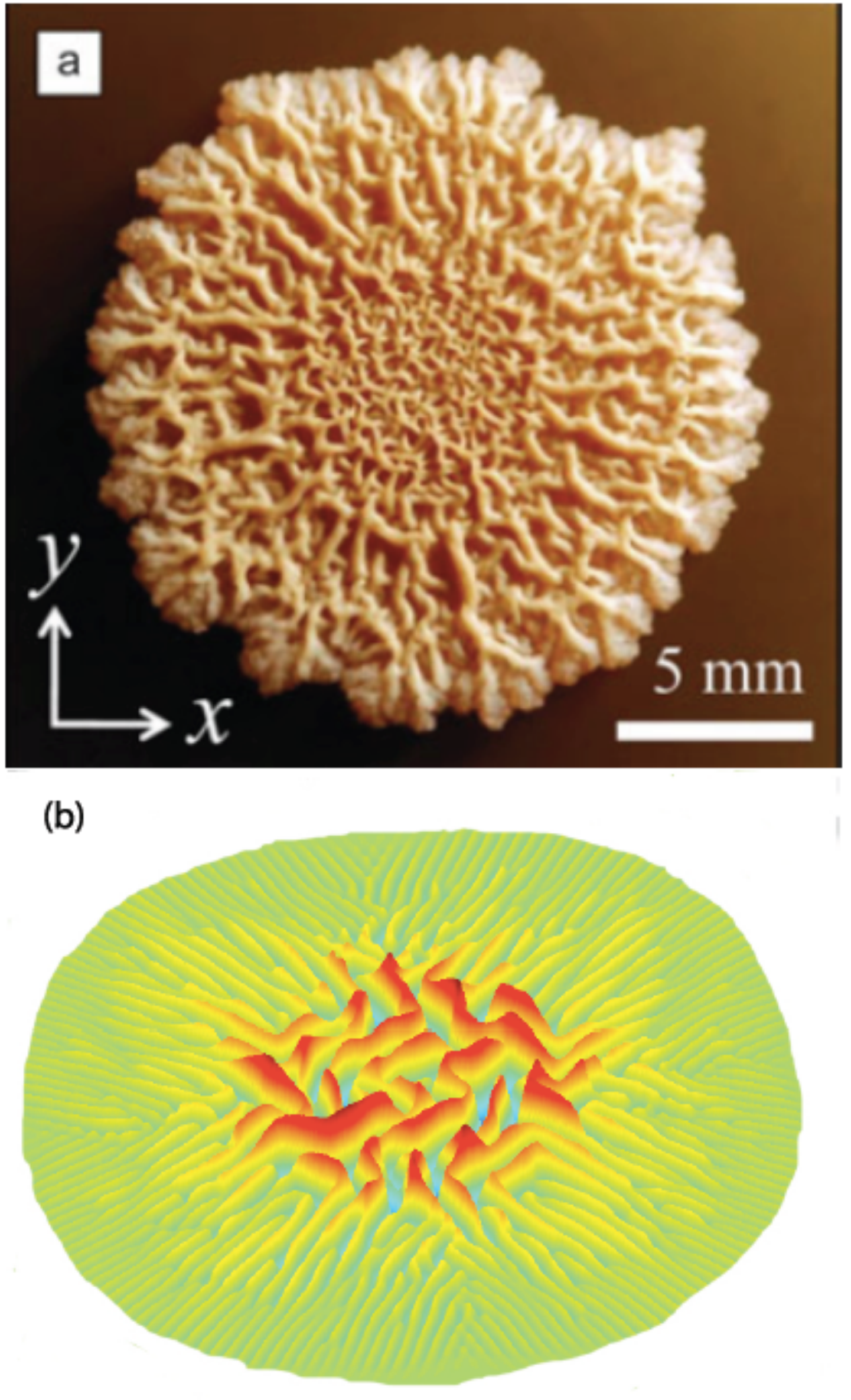}     
\caption{(Color online) 
(a) {\it B. subtilis} biofilm grown on the air-solid surface of agar gel containing water and nutrients. Reprinted with the permission of Cambridge University Press from Wilking JN, Angelini TE, Seminara A, Brenner MP, Weitz DA, MRS Bulletin,  36: 385-391 (2011)
\cite{osmotic}. (b) In silico film showing successive wrinkle branching.}
\label{fig2}
\end{figure}

Typical wild type {\em B. subtilis} biofilms on agar show a central core with highly connected wrinkles surrounded by a corona with radial branches, as in Figure \ref{fig1}. Successive branching is also reported, see  Figure \ref{fig2}. These structures differ from those commonly observed {in biofilm pellicles formed on air-liquid interfaces \cite{trejo}.  Such  pellicles spread on a space confined by the walls of the vessel containing the liquid.  Ref. \cite{trejo} assumes an expression for the compression stresses and performs a stability analysis of the equilibria of a plate model, concluding that a buckling instability should appear above a threshold. The presence of walls stresses the pellicles as they grow, producing wrinkles. In contrast, the biofilms studied here spread on an air-agar interface with no boundaries around. The mechanism for wrinkle formation is bound to be different, even if elastic deformations are involved too.  Ref. \cite{benamar3} uses a neo-Hookean energy to study the elastic deformations of a growing circular film moderately adhered to a rigid substrate and subject to anisotropic growth. Tracking bifurcations in families of analytical solutions as the anisotropy degree increases, they predict the appearance of contour undulations, delamination, buckling on the scale of the sample after delamination and, finally, a periodic distribution of folds.  As Ref. \cite{benamar3} points out, such buckling patterns are not experimentally observed in {\it B. subtilis} biofilms. However, the analysis may be useful to understand delamination and folding phenomena. Those models do not consider the microscopic processes occurring in a biofilm. The complex interaction with the substratum is reduced to attachment using a collection of springs. On the other hand, neither energy nor linear stability approaches  answer the question of how different patterns emerge and evolve from one to another. Here we aim to compute the development of wrinkled structures as the biofilm expands horizontally, starting from an almost flat initial state.  We will solve time dependent equations coupling the film to the substratum. The residual stresses will be computed numerically averaging information from growth and water absorption processes. This allows to observe how wrinkles develop and how they respond to localized variations in the microstructure. We recall below some experimental observations to bear in mind in the sequel.}

During biofilm expansion,  {\em B. subtilis} populations  differentiate in several types in response to local environments created by growth, waste production, nutrient consumption and cell-cell communication \cite{hera}. Cells produce a signaling molecule, ComX. When a threshold concentration is reached, a percentage of the population of cells expresses some genes that induce a change in their phenotype, becoming surfactin producers. Surfactin may signal other cells (that do not make surfactin) to produce extracellular matrix. As the biofilm thickens, starving cells in the top become inert spores.

Initially, part of the cells are motile and express flagella allowing
to swarm on the surface. As the production of ECM increases, most cells lose their individual motility \cite{hera}. Flagella are downregulated and biofilm spread is influenced by ECM production \cite{seminara}. In early stages of the biofilm evolution, dividing cells interact with their neighbors and push them. The biofilm becomes thicker and the edges spread slowly on the surface.  Later increase in the exopolysaccharide   concentration raises the osmotic pressure, causing swelling of the biofilm due to water intake from the agar gel \cite{seminara}.  Cells are pushed from the bulk outwards, speeding up horizontal growth and increasing nutrient intake from the agar surface.

The origin of the wrinkles seems to be controlled by the mechanical properties of the biofilm, which are in part governed by the production of extracellular matrix. In early stages of the evolution of the central biofilm core, localized cell death takes place at high density regions due to biochemical stress. Cell death is usually followed by cell shrinkage and resorption.
Lateral mechanical stresses built up during growth are relieved at those sites through vertical biofilm buckling, leading to wrinkle formation \cite{asally}. Wrinkles are filled with water extracted from agar, as in Figure \ref{fig3}, forming a network of channels that enhances nutrient and waste transport driven by surface evaporation \cite{wilking}.  {The emergence of the outer radial wrinkles shown in Fig. \ref{fig1}(a) or successive wrinkle branching in Fig. \ref{fig2}(a) is less understood. }

Recent developments in the study of the wrinkling of gels and thin films on viscoelastic substrata might have a counterpart in biofilm growth.  A difference with cellular films though, is that there is no feedback between mechanical stresses, rates of swelling and growth and cellular differentiation. Swelling of the outer biofilm corona might produce radial wrinkles provided the center of the biofilm was stiffer and swelled less than the corona. 
This ``corona instability" is commonly observed in swelling gels when they are radially graded \cite{mora}. Such gels do not grow but swell by absorbing water while their elastic modulus diminishes. Contact of a film with a viscoelastic substratum may also induce wrinkles, and select the dynamics that form them \cite{huang,liu}. Wrinkle patterns that switch smoothly from an intricate core to branching radial  wrinkles have been observed in thin films on a viscoelastic substratum due to solvent diffusion in the film, which creates residual stresses  and stiffness gradients \cite{ni}.

Based on the previous observations we propose a hybrid biofilm description, combining a stochastic treatment of cellular processes coupled to the continuum concentration fields and a continuum description of deformation mechanisms informed by cellular activities and the interaction with the substratum.
Biofilm growth and physiology rely on the diffusion and transport of nutrients, waste, and signaling molecules, facilitated by water.
The elastic deformation of the biofilm is described by  F\"oppl-Von K\'arm\'an equations with residual stresses generated by  growth and swelling  \cite{landau,benamar1,benamar2}, formulated for a thin biofilm expanding on a substratum \cite{huang,ni} containing water and nutrients.
We  {introduce a strategy to estimate} residual stresses, as well as spatial variations in the elastic parameters of the biofilm, averaging and smoothing stochastic information on cell type distribution, cell division and water absorption. 
We are able to reproduce qualitatively patterns and trends observed in experiments, gaining insight on the way biofilms are shaped. {We argue that stiffness gradients due to growth and swelling
may cause wrinkle branching and  wrinkled coronae.}
We identify governing parameters and study their influence on the wrinkled structure. The substratum properties should be carefully measured since they largely affect the time scales for biofilm deformation. Variations in the values of its viscosity and Poisson ratio may change the wrinkling time scales from minutes to days, as argued in Section \ref{sec:wrinkle}. Experimental observations show time variations too, compare Figures 1(a) in Refs. \cite{hera} and  \cite{wilking} after four and one days, respectively.  
{The influence of the agar substratum on the dynamics of biofilms and bacterial communities has already been noticed in different contexts. 
Ref. \cite {allen} shows the relevance of the agarose concentration in the transition from planar to three dimensional behavior during the invasion of a three dimensional agar matrix by {\em E. coli}. 
Ref. \cite{smets} observes that the water content of the agar substratum affects the motility of {\em Pseudomonas} strains.
}


Section \ref{sec:models} describes the basic structure of the hybrid model. Cellular processes are considered in Section \ref{sec:cell}. Stochastic division, death, differentiation and EPS production mechanisms are introduced. Section \ref{sec:shape} discusses film deformation according to F\"oppl-Von K\'arm\'an equations and presents water absorption strategies, together with procedures to estimate residual stresses and spatial variations in the elastic constants. 
The coupling between stochastic and continuous descriptions is detailed in Section \ref{sec:coupling}. Simulations of biofilm evolution activating single or a few mechanisms are presented throughout the paper. Section \ref{sec:results} explores the coupling between processes through a few selected tests. Finally, Section \ref{sec:discussion} summarizes our conclusions.

\section{Hybrid model}
\label{sec:models}

{To investigate the influence of processes at bacterium level, cells (or small clusters of similar cells) are represented by discrete `agents'. Different modeling frameworks describe those `agents' and their evolution in diverse ways. Cellular automata distribute them over the sites of a regular lattice \cite{glaziergas}. The `agents' have internal state variables, that change according to a set of rules, stochastic or deterministic. Standard cellular automata for biofilms in flows and in porous media allocate biomass (cells and EPS), fluids and solids over the lattice sites \cite{laspidoucellular,kapelloshierarchical}. The dynamics of the biomass in each tile is governed by rate equations coupled to concentrations and fluid equations. Whenever a threshold is surpassed, a neighboring tile is occupied, pushing biomass further in the direction of minimal mechanical resistance. Below a threshold, dead occurs with a certain probability. Biomass fragments can detach due to erosion by the flow. Refs. \cite{kapelloshierarchical,laspidoumechanical} use this approach to predict local diffusion coefficients, permeability coefficients and Young modulus variations taking into account the microstructure.

Cellular Potts approaches (in particular, the Glazier-Graner-Hogeweg variant) have been mostly applied to describe aggregation and migration of cells and tissue growth, see Ref. \cite{glazierpotts} for a review of applications.  These  models use energy based Monte-Carlo updating and are able to account for cell volume and shape. This is important when describing interactions strongly dependent on cell geometry \cite{glaziergas}. Effective (not physical) energies  typically include terms representing contact between neighboring cells, a volume constraint on cells, plus chemotaxis terms \cite{glazierpotts}.  Liquids and other substances are represented as additional cells. According to Ref. \cite{glaziergas}, the DAH (differential adhesion hypothesis) guarantees that final shapes correspond to the minima of effective energies. This hypothesis assumes that an aggregate of cells behaves as a mixture of immiscible fluids \cite{glaziergas}. 

Individual-based models developed for biofilms in aquatic media consider cells as particles (usually spheres) distributed in a three dimensional space, whose volume and mass grow due to nutrient consumption, see Refs. \cite{picioreanusurvey,picioreanurods} for a survey of applications. Cell division and death occur when specified size thresholds are reached.  Daughter cells are distributed randomly and dead cells are removed. The EPS matrix may be represented as a fluid continuum, particulates and/or capsular EPS. Biomass is eroded according to a law depending on shear. Agent growth, division, death and shrinking cause agent movement, via shoving algorithms and a pressure field. Biomass is advected following a Darcy's law involving the pressure and the dynamic viscosity of the biofilm, which is therefore assimilated to a fluid \cite{picioreanusurvey}. To investigate the effect of the shape of bacteria on the alignment of microbial communities expanding on a surface, rod-spring and sphere-spring representations have been recently implemented \cite{picioreanurods}. Each bacterial rod or sphere is characterized by its mass, velocity and position. The EPS matrix is represented by springs that join the rods together, and to the substratum. Cells divide when they reach a threshold mass due to nutrient consumption. New cells and springs must be created in the network. Then, the system relaxes to a new equilibrium following Newton's laws. A similar strategy is applied in Ref. \cite{allen} to describe the influence of the shape of bacteria in the invasion of a three dimensional agar matrix by {\em E. coli}.  

We are interested here in understanding the elastic deformations caused by cell division, death, EPS production, swelling, and the interaction with the substratum. A framework facilitating the coupling of the relevant microscopic and macroscopic phenomena is advisable. We resort to a cellular automata based hybrid description.}
Hybrid models combine stochastic representations of cellular processes with continuous descriptions of other relevant fields, in our case, concentrations and deformations. This allows to take into account the inherent randomness of individual bacterial behaviors observed  in the experiments \cite{hera,wingreenbacillus}. It also permits to consider local interactions around each cell to describe  differentiation processes. 

\begin{figure}
\centering
\includegraphics[width=9cm]{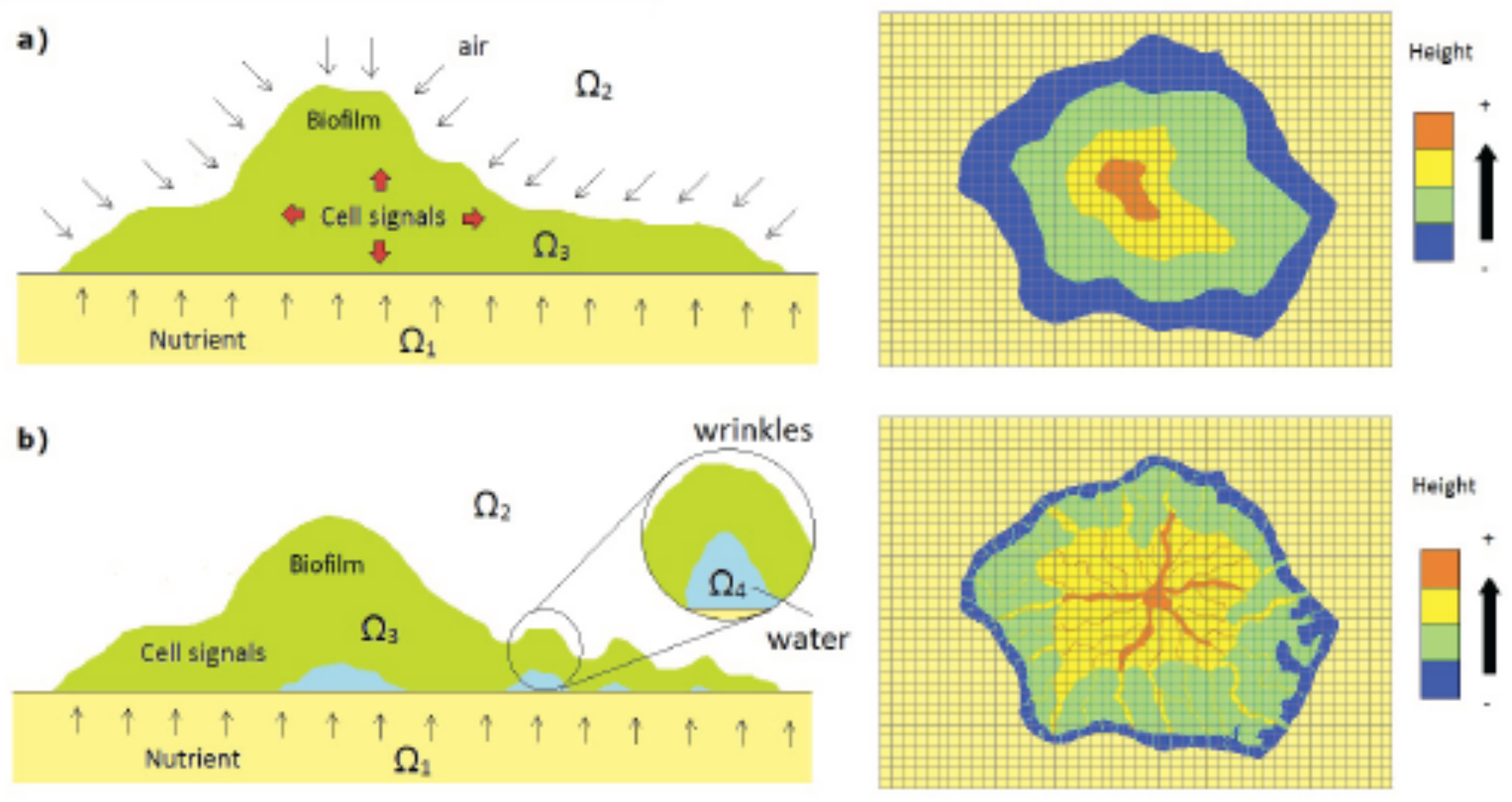}
\caption{(Color online)
Schematic representation of a biofilm.
(a)  Biofilm without water network, (b) Biofilm with water channels.
For computational purposes, space is partitioned in a grid of tiles of size
$a$, $a$ being the average length of a bacterium. Each tile may be filled
with air, water, agar, or biofilm biomass (either a bacterium or ECM). This 
grid is used to discretize the continuum concentration and displacement
fields, and to track the status of individual bacteria, which constitute
a fraction of the total biofilm biomass.}
\label{fig3}
\end{figure}

{In this model,} cells are regarded as creatures living in a grid, that reproduce, differentiate or die with certain probability, informed by the status of the concentration fields nearby. In a biofilm, cells are immersed in extracellular matrix. The dominating mass transport mechanism in the system is the diffusion produced by the presence of chemical concentration gradients of the chemical compounds (oxygen, nutrients, waste products and autoinducers) between the biofilm and the other phases (agar, water and air). Cellular processes inform inner mechanical stresses that shape the biofilm.

We consider a \textit{B. subtilis} biofilm patch (subdomain $\Omega_3$ in Figure \ref{fig3}(a)) 
attached to an agar-gel surface (subdomain $\Omega_1$) and exposed to the atmosphere 
(subdomain $\Omega_2$). 
Eventually, the biofilm may contain channels carrying water (subdomain $\Omega_4$) dragged from the agar substratum, as in Figure \ref{fig3}(b). To perform their metabolic activities, the biofilm obtains nutrients and other substances from the gel-water substratum, and from the atmosphere. All the subdomains are divided in a grid of cubic tiles. The spatial step $a$ chosen for this description was the size of one bacterium (around $3$-$5$ $\mu$m) to be able to consider individual interactions among bacteria.  Each tile can be filled with either agar, air, fluid or biofilm biomass (either cells or ECM).  Partition in tiles allows to track the status of individual bacteria, which represent a fraction of the total biomass, and discretize the equations governing the evolution of continuum fields.

\begin{figure}[!ht]
\centering
\includegraphics[width=9cm]{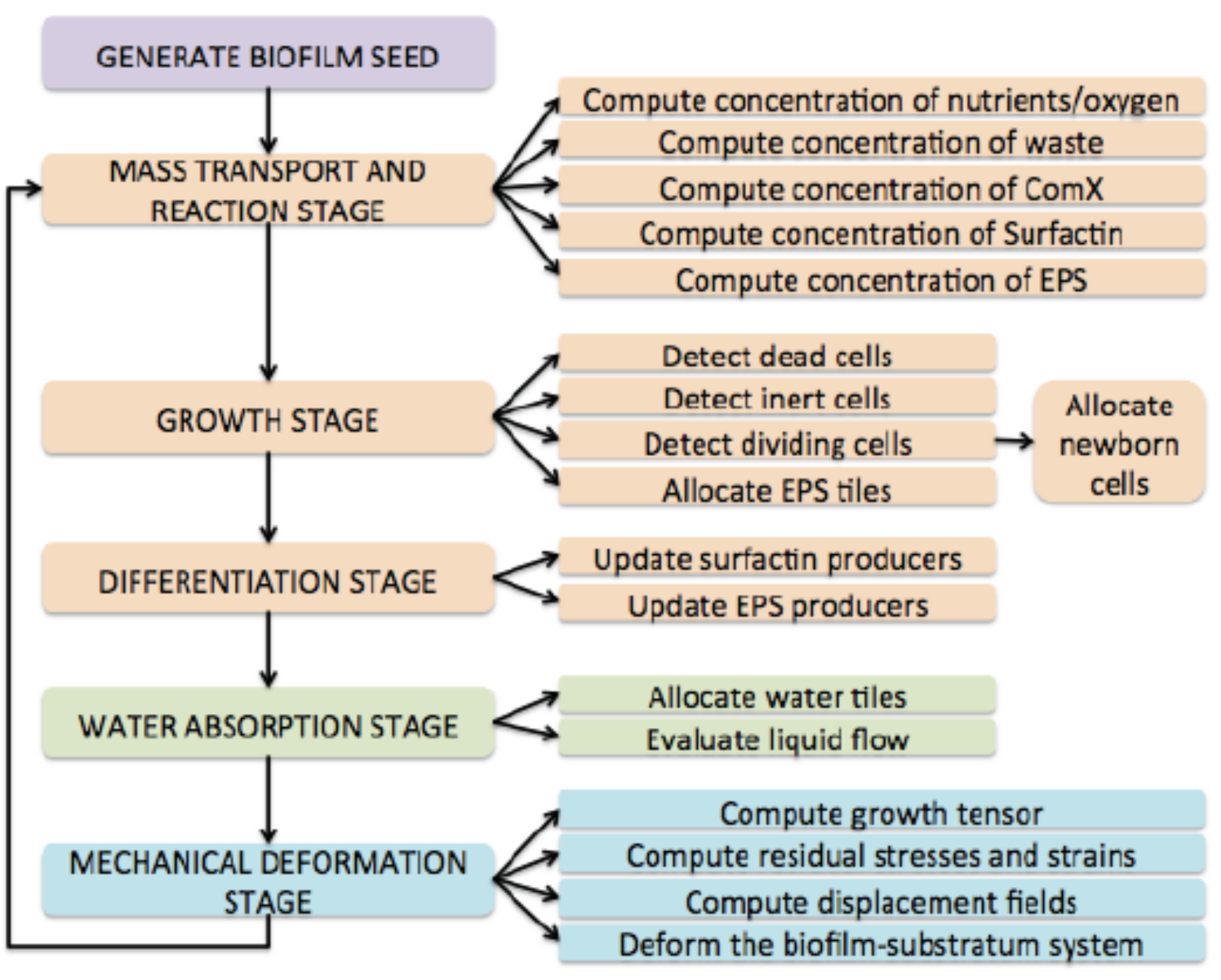}
\caption{(Color online) Flow chart of the hybrid model.}
\label{fig4}
\end{figure}

The computational evolution of a biofilm in our model is schematized in Figure \ref{fig4}. First, we update the cellular status, detecting dividing, dead or differentiated cells and allocating newborn cells and water. Then we deform the biofilm according to its internal stresses and revise the concentration fields and the cellular status. The next sections describe how we take into account cellular behavior and mechanical processes.

\section{Description of cellular processes}
\label{sec:cell}

It is commonly accepted that biofilms are able to generate molecular signals as cell-cell communication  mechanisms to activate gene expressions that unleash different survival strategies adapted to the environmental conditions \cite{purevdorj,hera}.
The basic approach relies on the idea that the same organism that is sensible to certain chemical inducers is also responsible of producing those compounds, creating a self-feeding loop which regulates both the genetic expression and the generation of autoinducer. Quorum sensing mechanisms provide a way for bacteria to count their population prior to unleashing different behaviors such as differentiation, virulence or spreading  \cite{fuqua,kreft,wingreenfeedback,wingreenintegration}. 

Differentiation is a mechanism through which some bacteria inside a population {display} physiological alterations, developing different phenotypes without a change in their genotype. {These modifications lead to the development of specialized cells with certain characteristics which perform specific tasks.}

\begin{figure}[!ht]
\centering
\includegraphics[width=8.5cm]{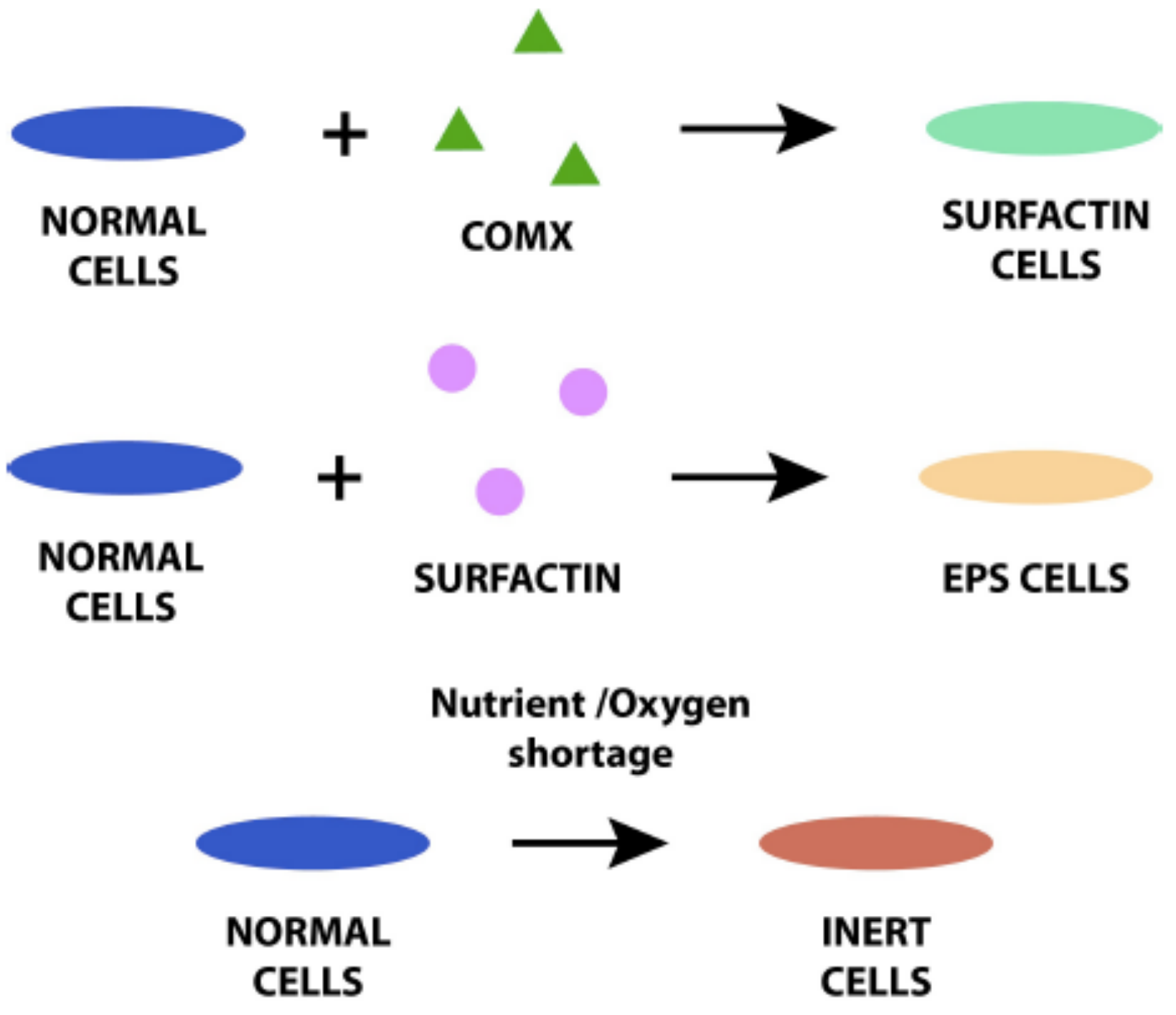}
\caption{(Color online) Directional signaling in {\it B. subtilis.} 
When a threshold concentration of ComX is reached, a percentage of the population of cells expresses genes that induce a change in their phenotype, becoming surfactin producers. 
Surfactin may signal other cells to produce extracellular matrix. Starving cells become inert  spores.}
\label{fig5}
\end{figure}

Differentiation processes have been observed to depend on the local conditions around each cell, in particular, the state of the neighbours and the local chemical composition found around them \cite{stewart,wingreenbacillus}. \textit{B. subtilis} provides a well known example, where initial biofilm cells may differentiate in four well defined cell types.  Two autoinducers, ComX and surfactin, relate the four types of differentiated cells \cite{hera}, see Figure \ref{fig5}. The different cell types are:\\
$\bullet$ {\it Normal cells.} They perform all metabolic activities and generate  ComX. They are sensible to all autoinducers and may be deactivated due to a lack of nutrients or oxygen. They may turn into surfactin generators or cells expressing matrix genes when threshold concentrations of  ComX or surfactin are reached.\\
$\bullet$ {\it Surfactin producers.} They generate surfactin and do not reproduce. Proliferation of these cells may be inhibited  by the presence of EPS producers in their neighborhood. \\
$\bullet$ {\it EPS matrix producers.} They secrete EPS matrix, and are able to reproduce at lower rate than normal cells because their metabolism also spends resources on generating  EPS. \\
$\bullet$ {\it Inert cells, or  spores.} These cells do not perform any metabolic activities, neither reproduction, nor differentiation. EPS producers become spores when there is a severe lack of nutrients  in the local environment of the cell.  Spores may  activate again when the environmental conditions improve.\\
As time progresses, a fraction of cells  expresses surfactin genes. Later on, an increasing number of cells secretes EPS matrix. Cells may also die as a result of biochemical stresses \cite{asally}. This has been shown to occur at the interface agar-biofilm as a result of biochemical stresses at localized high density regions. Eventually, sporulation starts  and few normal cells remain. Spores are usually located in upper regions of the biofilm, whereas normal cells are restricted to the bottom edges \cite{hera}. 

We describe below the rules for cell differentiation, death and reproduction.

\subsection{Cellular division, deactivation and death}
\label{sec:division}

When there is an excess of oxygen, the concentration of nutrients $c_n$ becomes the limiting concentration that restricts biofilm growth inside the  biofilm $\Omega_3$:
\begin{eqnarray}
c_{n,t} - D_{n,b}  \Delta c_n = - k_n {c_n\over c_n+K_n}. \label{cnb}
\end{eqnarray}
Nutrient uptake is simply described by a Monod law. Inside the agar substratum $\Omega_1$: 
\begin{eqnarray}
c_{n,t} - D_{n,a} \Delta c_n = 0.
\label{cna}
\end{eqnarray}
In the water phase $\Omega_4$ (if present):
\begin{eqnarray}
c_{n,t} - D_{n,l} \Delta c_n + {\bf v}\cdot \nabla c_n = 0. 
\label{cnl}
\end{eqnarray}
Here, $D_{n,b}$, $D_{n,a}$ and $D_{n,l}$ are diffusion constants in the biofilm, agar and water, respectively,  $K_n$ is the Monod half-saturation coefficient, and $k_n({\bf x})$ is the uptake rate of the nutrient at the particular location. It will be set equal to $k_n$ in tiles occupied by alive cells and equal to zero elsewhere. ${\bf v}$ is the transport velocity in the water channels.  On the interfaces agar-biofilm and water-biofilm (if present) we impose transmission boundary conditions (continuity of concentrations and fluxes). On the interface biofilm-air, the interface agar-air and the edges of the agar substratum, we have zero flux conditions. Initially, the nutrient concentration is usually taken to be constant $c_n=C$ in agar.

Tiles occupied by alive cells  ${\cal C}$ are assumed to divide with probability \cite{hermanovic}:
\begin{eqnarray}
P_d({\cal C}) = {c_n({\cal C}) \over c_n({\cal C}) +K_n}, \label{pdiv}
\end{eqnarray}
$c_n$ being the limiting concentration. Initially, we reallocate newborn cells inside biofilm by pushing existing cells in the direction of minimum mechanical resistance, that is, the shortest distance to the biofilm-air interface \cite{hermanovic,kapelloshierarchical}. 

\begin{figure}[!ht]
\centering
\includegraphics[width=8.5cm]{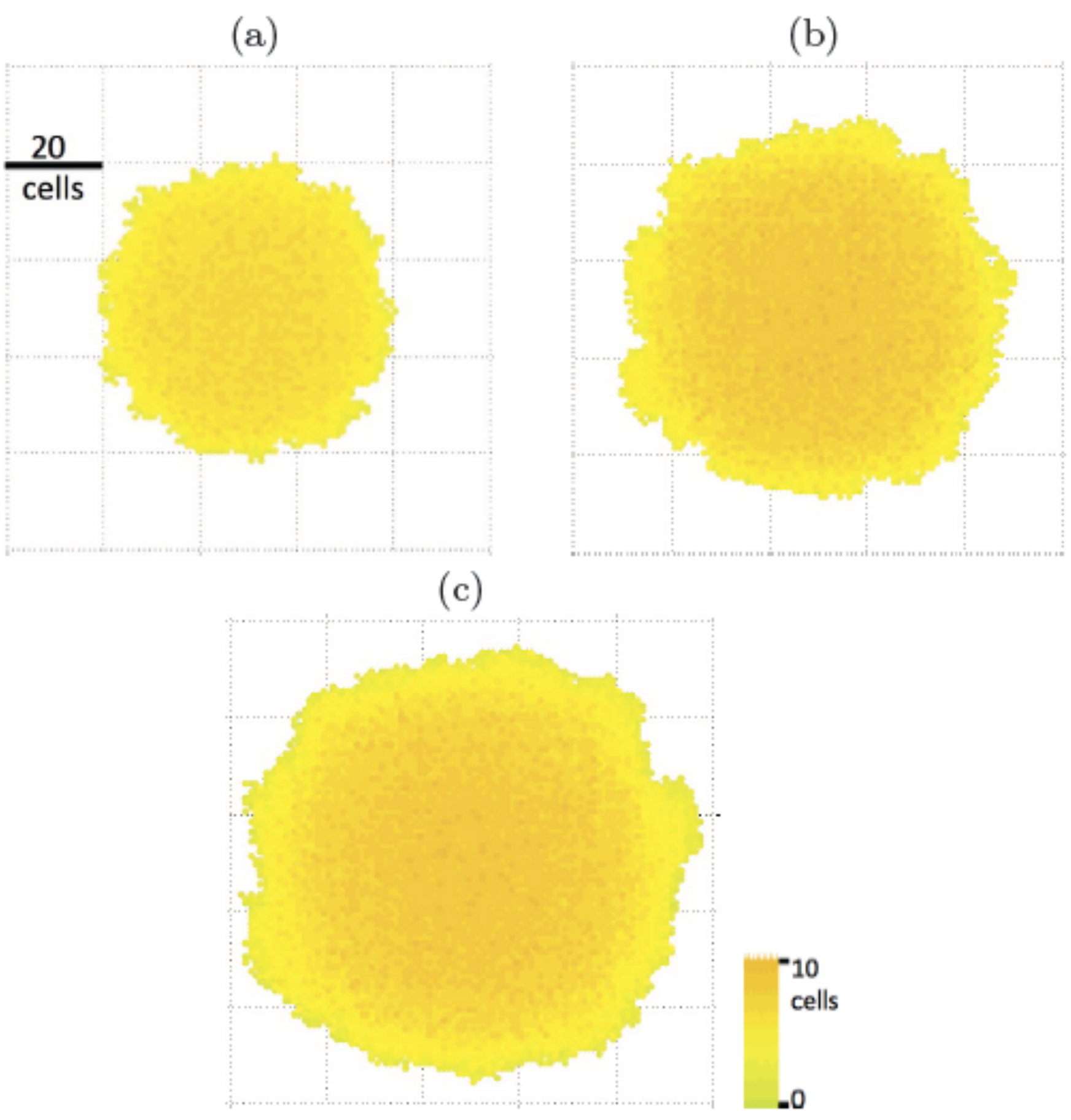} 
\caption{(Color online) 
Expansion of a circular biofilm seed driven only by nutrient consumption.
The initial seed is formed by two layers of cells with a diameter of $40$ cells. Only division and spread mechanisms are involved.  Snapshots taken every $100$ steps. The initial nutrient concentration in the substratum is $C=3 K_n$  and the ratio  ${k_n a^2 \over D_n K_n}= 8$. }
\label{fig6}
\end{figure}

Figure  \ref{fig6} illustrates the expansion of a biofilm seed. The initial round cluster grows in size in a radial way. Brick-like colonies are formed. Their border may advance in a wavy way depending on the velocity, as usual when a free boundary expands outwards \cite{hou}.
This spread mechanism may change as the biofilm develops and water is absorbed, as explained later.

The constants $K_n$ affecting the probability laws are numeric constants that should be calibrated with experimental data to have a real description in the simulations and do not have to match with saturation constants used in the source terms of the reaction-diffusion equations for the concentrations, which are related with a chemical saturation limit. 
To simplify, they have been chosen to agree with the values for saturation constants. 



Cells may suffer a shortage of nutrients, triggering a sporulation mechanism to preserve bacteria in a ``deactivated" state until the environmental conditions improve. To reflect this, cells with a local limiting concentration below a certain threshold  will become spores in our simulations.
We deactivate cells ${\cal C}$  with probability $1-P_d({\cal C})$ whenever $P_d({\cal C})< \gamma_d$  for a certain threshold $\gamma_d>0$. These cells may activate again when the conditions improve.

As mentioned before, cells may also die due to biochemical stress in high density regions where waste products and toxins accumulate but nutrients deplete. Taking for instance the concentration of waste $c_w$ at a location as an indicator of death, a cell ${\cal C}$ is scheduled to die with probability:
\begin{eqnarray}
P_w({\cal C}) = {c_w({\cal C}) \over c_w({\cal C}) +K_w}, \label{pdead}
\end{eqnarray}
whenever $P_w({\cal C})>\gamma_w$ for a certain threshold $\gamma_w>0$. The concentration of waste is governed by
\begin{eqnarray}
c_{w,t} - D_{w,b}  \Delta  c_w =  k_w  \sum_i \delta_{i,w}, \label{cwb}
\end{eqnarray}
inside the biofilm $\Omega_3$. 
$\delta_{i,w}({\bf x})$ equals $1$ in regions occupied by alive cells ${\cal C}_i$ and vanishes otherwise.
On agar-biofilm and air-biofilm interfaces we impose no flux boundary conditions. If a fluid phase $\Omega_4$ is present, we impose transmission boundary conditions at the interface and include a diffusion-convection equation similar to Eq. (\ref{cnl})
for waste transport in the liquid.
$D_{w,b}$, $D_{w,l}$ are the diffusion constants,  and  $k_w$  the production rate of waste and toxins at the particular location.

\subsection{Surfactin and EPS production}
\label{sec:surfactineps}

Normal cells may turn into surfactin generators when a threshold concentration of ComX is reached. ComX is a molecule related with quorum sensing mechanisms in \textit{B. subtilis} \cite{hera}. It is produced by all the cells in the biofilm, except those which are not metabolically active.  The concentration $c_{\rm cx} $ of ComX inside the biofilm $\Omega_3$ is governed by:
\begin{eqnarray}
c_{{\rm cx},t} -D_{\rm cx,b}\Delta c_{\rm cx} = k_{\rm cx}  
\left(1-{c_{\rm cx} \over c_{\rm cx} + K_{\rm cx} }\right)\sum_i \delta_{i,{\rm cx}},  \label{ccx}
\end{eqnarray}
with no flux boundary condition on agar-biofilm and air-biofilm interfaces. $k_{\rm cx}$ is the production rate, and $D_{\rm cx,b}$ the diffusivity. $\delta_{i,{\rm cx}}({\bf x})$ equals $1$ if cell ${\cal C}_i$ produces ComX  and vanishes otherwise. $k_{\rm cx}$ is multiplied by an inhibition factor for large concentrations.   A guess for $K_{\rm cx}$ may be the half-saturation constant of  ComX. If a liquid phase $\Omega_4$ is present
we impose transmission boundary conditions at the interface and include a convection-diffusion equation similar to Eq. (\ref{cnl}) for ComX transport in the liquid.
Normal cells ${\cal C}$ are assumed to become surfactin producers with probability:
\begin{eqnarray}
P_{\rm s}({\cal C}) = { c_{\rm cx}({\cal C}) \over c_{\rm cx}({\cal C}) +K_{\rm cx}}, \label{ps}
\end{eqnarray}
whenever $P_{\rm s}({\cal C})>\gamma_{\rm s}.$ The threshold $\gamma_{\rm s}$ reflects that differentiation should start when a minimum background concentration of ComX is reached.

When EPS producers are present, they act as inhibitors. We may set
$P_{s}=P_{inhib}\frac{c_{cx}}{c_{cx} + K_{cx}}$
with $P_{inhib}$ the ratio of neighbors producing EPS to total number of neighbors. If a cell is surrounded by  EPS producers, there is no any chance of differentiation into a surfactin generator. A similar effect is obtained setting a much lower threshold for differentiation into a surfactin producer compared to EPS producers, as we usually do in our simulations.

Surfactin acts as an autoinducer in normal bacteria, changing their phenotype to become EPS producers \cite{seminara}. The concentration $c_{\rm s} $ of surfactin inside the biofilm is governed by:
\begin{eqnarray}
c_{{\rm s},t} -D_{\rm s,b}\Delta c_{\rm s} = k_{\rm s}  
\left(1-{c_{\rm s} \over c_{\rm s} + K_{\rm s} }\right)\sum_i \delta_{i,{\rm s}},  \label{cs}
\end{eqnarray}
with no flux boundary condition on agar-biofilm and air-biofilm interfaces. $k_{\rm s}$ is the production rate, $D_{\rm s,b}$ the diffusivity and $K_{\rm s}$ the half-saturation of surfactin. $\delta_{i,{\rm s}}({\bf x})$ 
equals $1$ if cell ${\cal C}_i$ produces surfactin and vanishes otherwise.  When a liquid phase $\Omega_4$ is present transmission boundary conditions hold at the interface and  a convection-diffusion equation similar to Eq. (\ref{cnl}) governs surfactin transport  in the liquid.
We assume that active cells ${\cal C}$ not secreting surfactin become EPS producers with probability:
\begin{eqnarray}
P_{\rm e}({\cal C}) = \left( 1- { c_{n}({\cal C}) \over c_{n}({\cal C}) +K_{n}} \right) 
\left( { c_{\rm s}({\cal C}) \over c_{\rm s}({\cal C}) +K_s } \right), \label{peps}
\end{eqnarray}
whenever 
$ { c_{\rm s}({\cal C}) \over c_{\rm s}({\cal C}) +K_s }>\gamma_{e_1},$ 
which means that a minimum concentration of surfactin is required. The probability increases with the availability of surfactin and the scarceness of nutrients. This is consistent with measurements reported in Ref. \cite{zhang}. If we want to enforce a level of nutrient depletion to allow differentiation into EPS producers we may further impose $ { c_{\rm n}({\cal C}) \over c_{\rm n}({\cal C}) +K_n }< \gamma_{e_2}$. This may forbid differentiation below a certain height.

\begin{figure}[!htb]
\centering
\includegraphics[width=8.5cm]{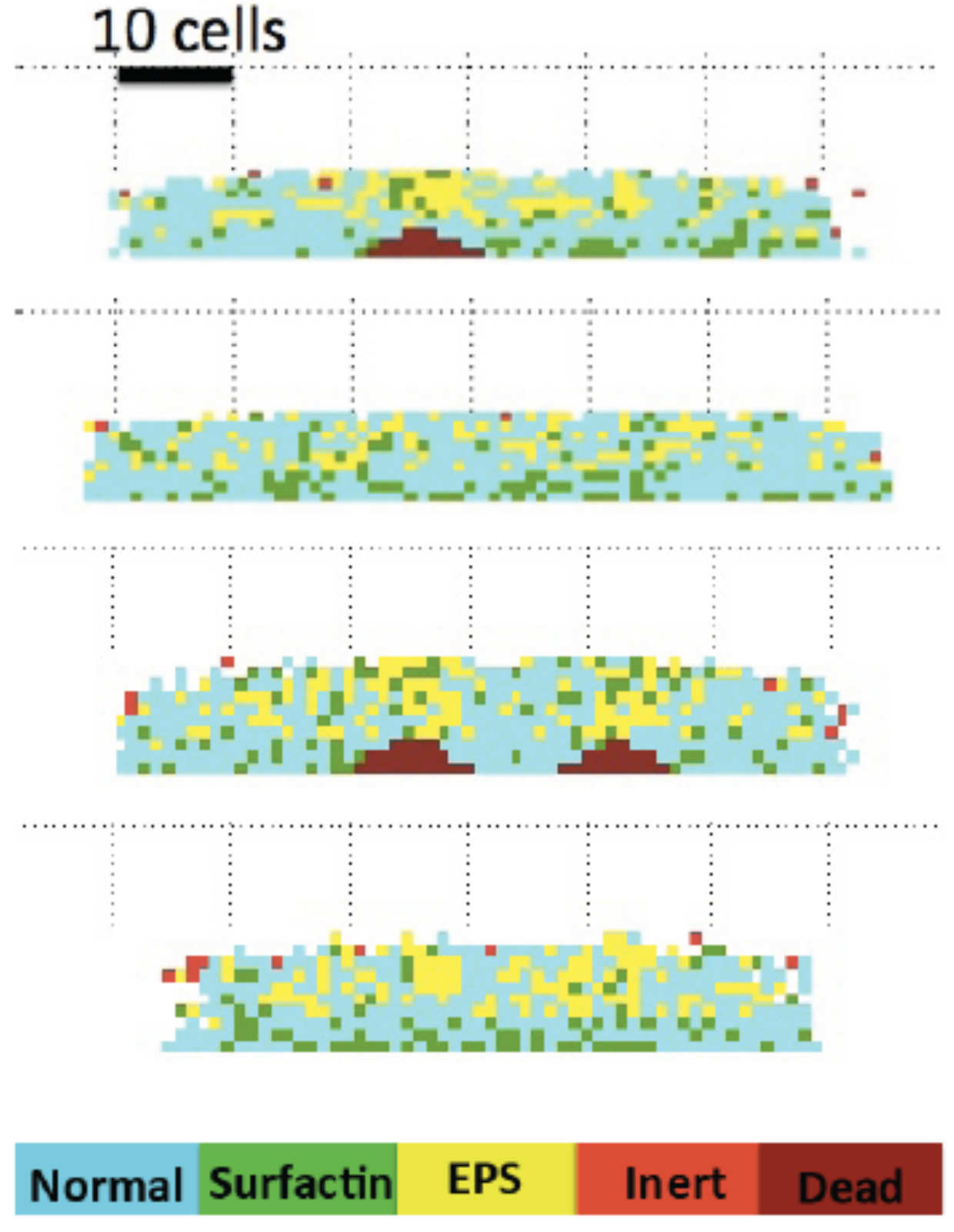}
\caption{(Color online)
Slices of a biofilm showing the cell type distribution during early stages of the differentiation process.  All the slices correspond to the same biofilm and are taken at the same time. The evolution started from an initial seed with a diameter of $60$ cells and variable thickness, represented by three small peaks. Dead cells appear first where the initial peaks where located. In later stages, inert cells proliferate at the top whereas normal cells are restricted to the bottom and the edges. 
Each colored box is one cell. Parameter values:  $C=K_n$, ${k_n a^2 \over D_n K_n}= 0.01$, ${k_w a^2 \over D_w K_w}= 0.001$, ${k_{cx} a^2 \over D_{cx} K_{cx}}=0.01$, ${k_s a^2 \over D_s K_s}= 0.8$, $\gamma_w=0.005$, $\gamma_d=0.00001$, $\gamma_s=0.001$, $\gamma_{e_1}=0.00001$,$\gamma_{e_2}=0.2$.
}
\label{fig7}
\end{figure}

Figure \ref{fig7} illustrates the distribution of cells when we activate the differentiation processes, in combination with standard growth and spread.
As the size of the biofilm increases, the concentration of ComX raises until the threshold    
$\gamma_s$ is reached. At this point bacteria start to differentiate and secrete surfactin. This triggers differentiation of cells into EPS producers when the surfactin threshold $\gamma_{e_1}$ is reached. As the number of EPS cells increases, the rate of differentiation into surfactin generators stabilizes and may even decay due to an inhibition mechanism induced by the EPS cells \cite{hera}. 
Relative proportions for each type of cell were adjusted by fitting parameters, but there should be an experimental calibration to improve these values.
Normal cells are found in the edges and near the interface biofilm-agar. EPS producers appear once surfactin is generated and a certain thickness is reached so that the nutrient concentration depletes. Dead cells concentrate on the bottom surface at inhomogeneity peaks when the initial seed has variable height, due to higher waste concentration. As the biofilm increases its thickness, scarcity of nutrients is felt in the upper part due to the high resistance to mass transfer across the increasing thickness. This effect induces the appearance of spores in the upper part, slowing down the growth of the biofilm. The geometric distribution of each type of cell seems to follow qualitatively trends reported in Refs. \cite{hera,zhang,asally}. 

{Implementing the differentiation strategy described above requires measuring a number of kinetic constants for different concentrations. Since these values are currently unknown, we have scaled the parameters to obtain the expected behaviors in small clusters of cells,  lowering the computational cost.
For Gram negative {\em Vibrio harveyi}, quorum sensing processes triggering bioluminescence  have been modeled in terms of feedback loops \cite{wingreenfeedback} and applying information theory to the quorum sensing circuit \cite{wingreeninformation}, which may reduce the number of parameters involved. Signaling parameters have been deduced from  in vivo data for {\em V. harveyi} \cite{wingreendata}.}

{For simplicity, the biofilm in Figure \ref{fig7} does not include the ECM phase, only cells. We might include it following the procedure described below.}
Once EPS production starts, a new concentration $c_e$ diffuses in the environment:
\begin{eqnarray}
c_{{\rm e},t} -D_{\rm e,b}\Delta c_{\rm e} = k_{\rm e}  
\left(1-{c_{\rm e} \over c_{\rm e} + K_{\rm e} }\right)\sum_i \delta_{i,{\rm e}}.  \label{ce}
\end{eqnarray}
$k_{\rm e}$ is the production rate, $D_{\rm e,b}$ the diffusivity and $K_{\rm e}$ the half-saturation of EPS. $\delta_{i,{\rm e}}({\bf x})$ equals $1$ if cell ${\cal C}_i$ is an EPS producer and vanishes otherwise. Initially, these substances are just compounds diffusing in the environment. As the concentration increases they become a polymeric envelop for the cells. 
This description can be improved taking into account the presence of molecules of different size.
Monomers diffuse, but longer molecules stay where they are produced. The concentration of momoners is governed by a diffusion equation coupled to the concentration of larger molecules, whereas the latter is governed by rate equations, that substitute Eq. (\ref{ce}).

The volume fraction of biofilm occupied by extracellular substances varies largely depending on the type of biofilm. Images in Ref. \cite{stonepnas} show biofilms in flows, formed by scattered cells embedded in a dominant EPS phase. On the contrary, imaging biofilms on air-semisolid interfaces at hight resolution shows a structure of densely packed cells glued by EPS, see references \cite{seminara}, \cite{hera} and \cite{asally}. Space is mostly occupied by cells. This suggests skipping the creation of an EPS phase and  thinking of it as a virtual glue that keeps cells together and modifies the mechanical properties of the biofilm, in particular, its elastic constants and the osmotic pressure.
If we choose to create an EPS phase, we may exploit the concentration $c_{\rm e}$ of EPS to generate new biofilm tiles filled with it. At any tile, we set
\begin{eqnarray}
P_{\rm m}({\cal C}) = { c_{\rm e}({\cal C}) \over c_{\rm e}({\cal C}) +K_{\rm e}}. \label{pm}
\end{eqnarray}
If this value is larger that a threshold $\gamma_m$, we create an EPS tile at that location and shift biomass around in the direction of least mechanical resistance. $\gamma_m$ should be calibrated so that the fractions of biomass
corresponding to cells and EPS fit the experimental observations.  The true cells are a fraction of the total amount of tiles occupied by biomass. They divide and differentiate according to the same probabilistic rules we have already described.

{Individual-based models for biofilms on agar surfaces represent the EPS as springs
joining the cells \cite{picioreanurods}. For biofilms in flows, expressing larger fractions of
EPS, each `individual' contains cellular mass and capsular EPS, that vary according to user defined rate equations. Additional EPS particulates may be created as EPS production
increases following user defined rules \cite{picioreanusurvey}. 
Refs. \cite{xavierupwards,xaviervibrio} study EPS dynamics in a quorum sensing framework, defining competing strains: EPS producers, EPS non producers, strains that downregulate EPS production for 
large cell densities. This is not the reported quorum sensing circuit for {\em B. subtilis}, see Ref. \cite{hera}.
Cellular automata models  for biofilms in flows distribute cellular mass and EPS in each lattice site following phenomenological rate equationsÊ \cite{laspidoucellular,kapelloshierarchical}. Spread to neighboring sites occurs as a threshold mass is surpassed. 
In both frameworks, cells are scheduled to die when their mass or volume falls below a threshold. 
No memory of their former presence, relevant for elastic deformations, is kept. 
Neither the relevance of waste accumulation for cell death, pointed out in Ref. \cite{asally}, nor the currently known quorum sensing circuits for EPS production are incorporated in  those models. }
 {Dynamic energy budget descriptions of cell dynamics \cite{birnir} might provide a  sophisphicated framework to describe the evolution of individual cells taking into account their mass, volume, maintenance, substance secretion and the presence of nutrients, oxygen and toxicants. However, extensions to biofilms are yet to be devised.}

\subsection{Nondimensionalization of the concentration equations}
\label{sec:dimensionless}

Nondimensionalization of the equations for the concentrations listed above reveals a key governing parameter for each concentration: $F={k a^2 \over K D_b}$, where $k$ is the uptake or production rate, $a$ the bacterium size, $K$ the half-saturation constant and $D_b$ the diffusivity inside the biofilm. Setting $\hat{c}={c \over K}$ and
$\hat{\bf x}={x\over a}$, the equations for concentrations become:
\begin{equation}
 -  \Delta_{\hat{\bf x}}  \hat{c} =  F\; f\Big({\hat{c} \over \hat{c}+1}\Big),
 \label{dimensionlessreaction}
\end{equation}
inside the biofilm, where $f$ represents the production or consumption terms, and
\begin{equation}
 - \Delta_{\hat{\bf x}}  \hat{c} + \hat{\bf v} \cdot \nabla_{\hat{\bf x}} \hat{c} = 0, \label{dimensionlesstransport}
\end{equation}
in the water channels, where $\hat{\bf v}={{\bf v} \, a \over D_{\ell}}$. We have suppressed the time dependence because diffusion  of chemical species is faster than the cellular processes considered. For each biofilm configuration, we solve these nondimensionalized stationary equations with the corresponding boundary conditions by a relaxation technique. Then we compute all the probabilities for division, death and differentiation, detect the different types of cells and allocate newborn cells. This generates a new biofilm configuration for which the procedure is repeated. A complete reproductive cycle of a bacterium was set as the basic time step for the cellular processes in the model (which is typically in the order of $26$-$45$ minutes depending on nutrient and temperature). This choice implies that our system is discrete in space and time \cite{hermanovic}.

\section{Description of mechanical processes}
\label{sec:shape}

Differentiated cells produce chemicals, which may induce further differentiation of other cells or trigger mechanical processes that alter the biofilm structure, improving its chances of spread and survival. EPS production increases the stiffness of the biofilm \cite{asally}, which may be identified with a  material undergoing macroscopic deformations caused by growth \cite{trejo} over a substratum. The spatial distribution of dead and inert cells,  together with water absorption,  may induce additional stiffness gradients. This motivates allowing our biofilm to move along the tiles according to microscopically informed continuum descriptions of the underlying mechanical processes. 

\subsection{Out-of-plane deformations and wrinkle formation}
\label{sec:wrinkle}

Due to the interaction of components in the extracellular matrix, dissipative phenomena take place inside soft tissues during deformation processes and the mechanical response of tissues if often viscoelastic. Nevertheless, growth induced deformations on time scales much larger than those of relaxation may be regarded as purely elastic \cite{benamar2}.
Since the height of real biofilms is about $10$-$100$ times smaller than their radius and they are initially thin and flat, we may use F\"oppl-von K\'arm\'an \cite{landau} equations for thin elastic plates to study their elastic deformation reducing dimensionality \cite{trejo}.  Residual stresses due to growth are incorporated in the description in terms of the growth tensor, see Refs. \cite{benamar1,benamar2}. The shape of the body is then determined by both growth and elastic deformation.  
The equilibrium deformation of a growing tissue is described by a variant of the F\"oppl-Von K\'arm\'an equations  \cite{benamar2}:
\begin{eqnarray}
D (\Delta^2 \xi - \Delta C_M)- h {\partial\over \partial x_{\beta}} \left( \sigma_{\alpha,\beta} 
{\partial \xi \over \partial x_{\alpha}}\right) = P, \label{plategrowth1} \\
{\partial \sigma_{\alpha,\beta} \over \partial x_{\beta}} =0, \label{plategrowth2}  
\end{eqnarray}
where $\alpha,\beta$ stand for $x,y$ and summation over repeated indexes is intended. The first equation describes out-of-plane bending 
$\xi(x,y)$ and the second one in-plane stretching for the displacements 
${\bf u}=(u_x(x,y),u_y(x,y))$. $(x,y)$ vary along the 2D projection of the 3D biofilm structure.
The bending stiffness is $D={E h^3\over 12(1-\nu^2)}$, and $h$ the initial plate height.
$P$ is the external pressure at the border of the sample. 
$C_M= {\partial (g_{zx}+g_{xz}) \over \partial x} + {\partial (g_{zy}+g_{yz}) \over \partial y}$
is a residual growth term. The components of the tensor $g$ are the growth rates, that is, the rate of volume supply per unit volume. 
Defining consistently a growth tensor $g$ is a non trivial issue. A possibility is to set $g=\nabla {\bf w}$  \cite{hejnowicz}, ${\bf w}$ being the vector field of displacement 
of growing biomass in the biofilm ${\bf w}$, which is determined by the cellular division and spread mechanism and  water absorption processes. 

Stresses $\sigma$ and strains $\varepsilon$ are defined in terms of in-plane displacements 
${\bf u}=(u_x,u_y)$ \cite{mora,benamar2} by:
\begin{eqnarray}
\varepsilon_{\alpha,\beta}={1\over 2} \left( {\partial u_{\alpha} \over \partial x_{\beta}} 
+ {\partial u_{\beta} \over \partial x_{\alpha}}  
+ {\partial \xi \over \partial x_{\alpha}} {\partial \xi \over \partial x_{\beta}}
\right) + \varepsilon_{\alpha,\beta}^0, 
\label{strain}\\
\sigma_{xx}= {E \over 1-\nu^2} (\varepsilon_{xx}+ \nu \varepsilon_{yy}), \;
\sigma_{xy}= {E \over 1+\nu} \varepsilon_{xy}, \label{stress}  \\
\sigma_{yy}= {E \over 1-\nu^2} (\varepsilon_{yy}+ \nu \varepsilon_{xx}). \nonumber
\end{eqnarray}
The Poisson ratio of different tissues has been measured to be about $0.49999$ \cite{tissue}.
We will set the Poisson ratio of the biofilm $\nu={1\over 2}$ (incompressibility) \cite{benamar2}. 
The average elastic modulus $E$ of a wild type {\em B. subtilis} biofilm has been measured to be about $25$ $kPa$ \cite{asally}. 
When the residual strains $\varepsilon_{\alpha,\beta}^0$ are due to growth, they are expressed in terms of the growth tensor as:
\begin{eqnarray}
\varepsilon_{\alpha,\beta}^0= -{1\over 2} \left( g_{\alpha \beta} + g_{\beta \alpha} 
+ g_{z \alpha} g_{z \beta} \right).
\label{residualstress}
\end{eqnarray}
A growing biofilm is in a state of compression due to cell division and, eventually, water absorption. Alternatively, this may be represented by a residual strain 
\begin{equation}
\varepsilon^0_{\alpha,\beta}({\mathbf x},t) = -\varepsilon_0({\mathbf x},t) \delta_{\alpha,\beta},  \quad \varepsilon_0>0.
\label{residual} 
\end{equation}
If we assume that cells do not grow at expense of their neighbors, $\delta_{\alpha,\beta}$ is a diagonal unit tensor in polar coordinates in a circular film.  


For a biofilm growing on a surface no external loads act on the edges, therefore, $P=0$. However, the interaction with the surface affects the deformation process and has to be included in the description.  
A correction to Eqs. (\ref{plategrowth1})-(\ref{plategrowth2}) applicable to thin elastic films growing on a viscoelastic substratum is proposed in Ref. \cite{huang}:
\begin{eqnarray}
{\partial \xi \over \partial t} &=& {1 - 2 \nu_v \over 2 (1-\nu_v)} {h_v \over \eta_v}
\Bigg[ D (- \Delta^2 \xi + \Delta C_M) \nonumber \\
&+& h {\partial\over \partial x_{\beta}} \left( \sigma_{\alpha,\beta}({\bf u}) 
{\partial \xi \over \partial x_{\alpha}}\right) \Bigg]
-{\mu_v \over \eta_v} \xi, \label{plategrowth1bis} \\
{\partial {\bf u} \over \partial t} &=& {h_v h \over \eta_v} 
\nabla \cdot {\bf \sigma({\bf u}) } - {\mu_v \over \eta_v} {\bf u}, \label{plategrowth2bis}  
\end{eqnarray}
where $h_v$ is the thickness of the viscoelastic substratum and $\mu_v$, $\nu_v$, $\eta_v$  its rubbery modulus, Poisson ratio, and  viscosity, respectively.

The equilibrium equations describe possible equilibrium configurations, and yield information on their stability or changes of stability. To describe the dynamics of wrinkle formation and the finally selected pattern we must solve the time dependent equations.  Equations (\ref{plategrowth1bis})-(\ref{plategrowth2bis}) were derived for the interface between the thin film and the substratum.
This interface moves vertically following the displacement field $\xi$. This shifts all the tiles in the three dimensional biofilm-substratum system vertically.

Choosing new dimensionless variables:
\begin{eqnarray*}
\hat {\bf x}= {{\bf x} \over L}, \quad \hat {\bf u}= {{\bf u} \over L}, \quad
\hat \xi= {\xi \over h}, \quad \hat \sigma={\sigma \over E}, \quad \hat t ={t \over T},
\end{eqnarray*}
and setting $L=\gamma h$, the dimensionless equations read:
\begin{eqnarray}
{\partial \hat \xi \over \partial \hat t} &=& 
\Bigg[  {12(1-\nu^2) \gamma^2}  {\partial\over \partial \hat x_{\beta}}
 \left( \hat \sigma_{\alpha,\beta}(\hat{\bf u}) {\partial \hat \xi \over \partial \hat x_\alpha}\right) 
 \nonumber \\
 &+&  (- \Delta_{\hat{\bf x}}^2 \hat \xi + \Delta_{\hat{\bf x}} \hat C_M) \Bigg]  
- T {\mu_v \over \eta_v} \hat \xi, \label{plategrowth1adim} \\
{\partial \hat{\bf u} \over \partial \hat t} &=& \tau \;
\nabla_{\hat{\bf x}} \cdot {\bf \hat \sigma(\hat{\bf u}) } - T {\mu_v \over \eta_v} \hat {\bf u}, \label{plategrowth2adim} 
\end{eqnarray}
{with 
\begin{eqnarray}
T &=& {2 (1-\nu_v) \over 1-2 \nu_v}{\eta_v h \over h_v} {12(1-\nu^2) \gamma^4\over E }, 
\nonumber \\
\tau &=& 24 {(1-\nu_v)\over (1-2\nu_v) } (1-\nu^2)\gamma^2.
\label{timesdeformation}
\end{eqnarray}  
We set $L$ equal to the maximum averaged radius of the biofilm in study. 
Common experimental values for the parameters defining the spatial scales are $h\sim 100 \mu$m,   $h_v \sim 100 h$. The size of the biofilm may vary with the nutrients and the agar gel nature. Figure 1(a) in Ref. \cite{hera} shows a wrinkled biofilm with $L \sim 50 h$ grown in four days on a $1.5\%$  agar surface. Figure 1 (a) in Ref. \cite{wilking} reproduces wrinkled biofilms with $L \sim 50 h$ within one day. 

The linear terms  $T {\mu_v \over  \eta_v} \hat {\bf u}$ and  $T {\mu_v \over \eta_v} \hat \xi $ are just damping contributions from the substratum slowing down the evolution. We will neglect them in our preliminary studies.
Wrinkling behavior is observed when the coefficient $12(1-\nu^2) \gamma^2$ is large enough, so that the contribution of nonlinear effects and residual stresses in Eq. (\ref{plategrowth1adim}) becomes relevant and drives the plate out of the flat equilibrium state, see Figure \ref{fig8}.

\begin{figure}[!ht]
\centering
\includegraphics[width=8.5cm]{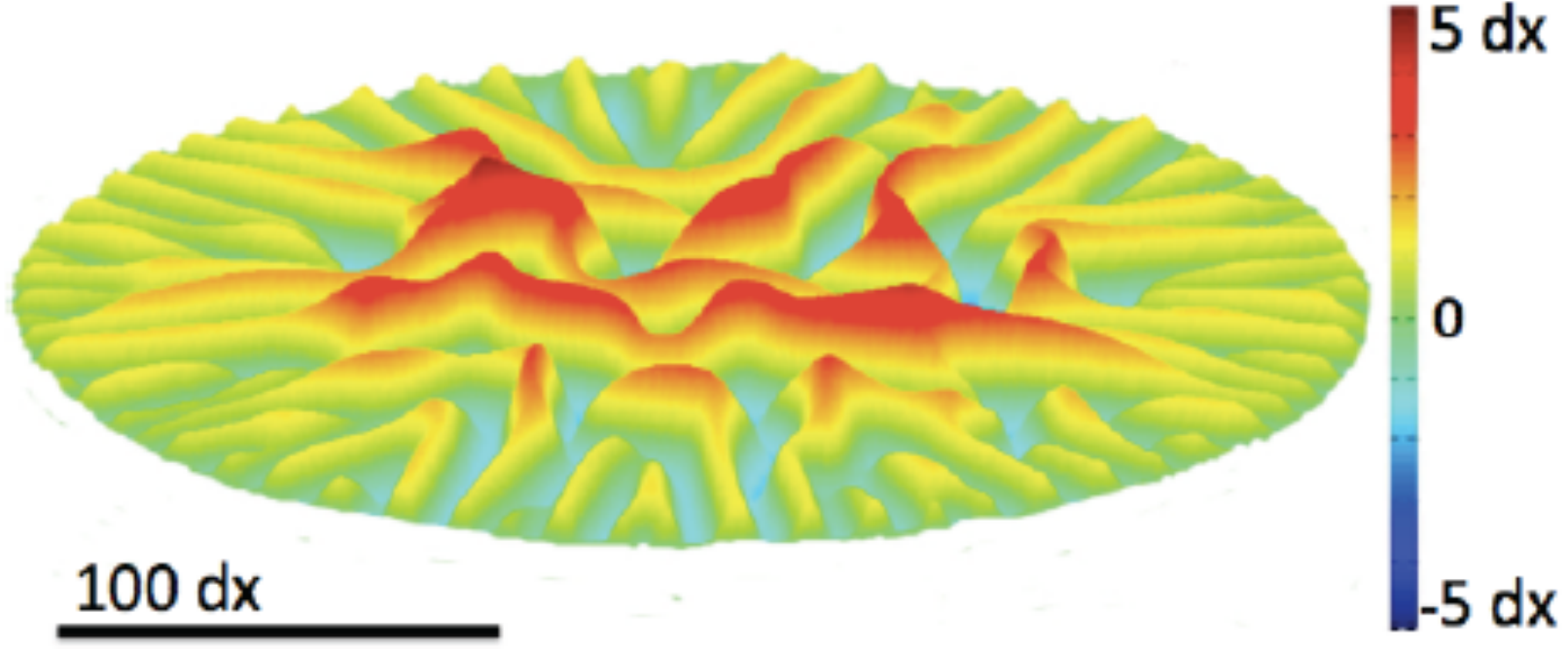} 
\caption{(Color online)
Connected network of wrinkles switching from an intricate core to radial branches, formed in a homogeneous circular film when a circular front of residual compression stresses of magnitude $\varepsilon_0=0.1$ expands at constant velocity. 
Parameter values: $\nu=0.5$, $E=25\, kPa$, $\nu_v=0.45$, $\mu_{v}=0$ and $\gamma=16.$ $dx=0.1\,h$, where $h$ is the film thickness before wrinkling.}
\label{fig8}
\end{figure}

The factor $\tau$ in Eq. (\ref{plategrowth2adim}) separates the time scales for the evolution of ${\bf u}$ and $\xi$. The closer to $0.5$ the Poisson ratio $\nu_v$ is, the larger this separation is. The in-plane displacements ${\bf u}$ may reach a quasistatic situation in which their evolution is driven by slower changes in the residual stresses created by growth, swelling, and off-plane displacements.

The constitutive parameters for an agar gel may vary depending on the agar source, its content of agar and other  products, in particular, water. When an hydrogel is fully swollen, its mechanical behavior is similar to those of rubber-like materials \cite{hydrogel}, which have a Poisson ratio of about 0.5. For some  agar gelatines the Young  modulus is measured to be $0.4999$, with an elastic modulus about $27$ kPa \cite{tissue}. In other agar gels these numbers  go down to $\nu_v=0.32$ with an elastic modulus  $E_v=52$ kPa \cite{agargel}. This modulus increases with the agar concentration from $\sim 30$ kPa in $0.5\%$ samples up to $700$ kPa in $5\%$ samples. The viscosity $\eta_v$ of agar gels is reported in the range of $100$ cp = $0.1$ Pa$\cdot$s  for $1.5\%$ samples, see Ref. \cite{seaweed}. As the agarose concentration increases to $8\%$, the viscosity raises to $1-1.5$ Pa$\cdot$s \cite{agartemp}. These values increase as the temperature diminishes. According to Ref. \cite{agartemp}, no measurements can be obtained below $36^o$ C.  Standard experiments are carried out at room temperature, usually below that threshold. Effective values for viscosities in a MPa$\cdot$s range are adjusted in Ref. \cite{eviscosity}.

A large variability in the time scales may arise due to uncertainty in the experimental values of $\nu_v$ and $\eta_v$. Setting for instance $\nu_v=0.4999$, $\eta_v=0.1$ Pa$\cdot$s, we find $T\sim 31 $ hours. This value varies enormously depending on $\nu_v$, that changes with the water content. Switching to $\nu_v=0.45$ and $\eta_v=1$ Pa$\cdot$s, we find $T\sim 2480 $ s $\sim 40$ m.  The time scale scale for ${\bf u}$ is $T'={T \over \tau}$ ranges from $10^{-3}$ s in the first case to $10^{-2}$ s in the second one.  If $\eta_v$ enters the range of kPa$\cdot$s or MPa$\cdot$s, this value increases by a factor $10^3-10^6$.

\subsection{Water absorption and water channels}
\label{sec:water}

As mentioned earlier, variations in the osmotic pressure extract water from the agar substratum. In Ref. \cite{seminara}, a macroscopic model for water migration from the agar substratum into the biofilm is proposed in terms of the volume fractions of water and biofilm at each location. The biofilm is formed by cells and extracellular matrix, which is often considered a gel with ability to swell. As observed in Ref. \cite{wilking}, flat regions of the biofilm are very resistant to flow. Water cannot be driven inside without fracturing the biofilm or delaminating it from the substratum. By contrast, water flows beneath the wrinkles, forming an intricate network of channels \cite{wilking}. Wrinkles origin in the core of the biofilm associated with dead areas \cite{asally}. 
Water flows along the channels driven by surface evaporation \cite{wilking}.

These observations suggest a strategy to incorporate water into our biofilm description.
Water absorption by the biofilm increases the volume of the cells and the EPS phase. Enlarging the size of the tiles to reflect that fact is unpractical. We resort instead to inserting water tiles in the biofilm, representing water absorption by the biomass. In that way, the biofilm contains a volume fraction of adsorbed water that changes its volume. A low cost strategy, easy to implement in our hybrid framework is the following. As in Ref. \cite{seminara}, we propose for the osmotic pressure a law of the form $\pi = \Pi \phi$, but replacing the volume fraction  $\phi$ of biomass by our information of the biomass available at biofilm columns 
\begin{equation}
\pi = \Pi {N \over N_{max}}, 
\label{pressure}
\end{equation}
where $N$ is the number of tiles occupied by biomass in a column and $N_{max}$ is the total number of tiles in the column, including water. Water tiles are created with probability:
\begin{equation}
P_l({\cal C}) = { \pi({\cal C}) \over \pi({\cal C}) + E({\cal C})},
\label{pwater}
\end{equation}
where $E({\cal C})$ is the Young modulus at tile ${\cal C}$ and the height of the columns in Eq. (\ref{pressure}) is updated as we create water tiles. The status of neighboring tiles is shifted in the direction of minimal mechanical resistance, except when water occupies a dead cell. More water will be absorbed in the columns containing a larger fraction of biomass, and in softer regions.   Scattered inner tiles do not constitute a separate phase, but are considered to be part of the swollen biomass. 

We compute $E({\cal C})$ modulating the measured reference value of the Young modulus of the biofilm with an average of weights representing status of the neighboring tiles. Weights range from zero for water, a small positive value for dead cells, a larger value for normal cells, up to one for EPS producers. Alive cells attached to the substratum contribute much larger weights than the rest. $E({\cal C})$ will be small wherever we have dead cells. As Figure \ref{fig9} shows, those regions fill with water in successive steps, becoming a water phase that expands following the wrinkles.   Combining water transport through those channels, with the cellular processes described in Section \ref{sec:cell}, we see that cell death due to waste accumulation is reduced and the colony is able to expand in a sustained way. 
In early stages we do not allow newborn cells into this water phase due to the pressure gradients that drift water from the agar gel into the biofilm. As the water content of the agar gel diminishes we might allow cell expansion around the channels, as observed in Ref. \cite{wilking}. This would keep water inside the biofilm.

\begin{figure}[!ht]
\centering
\includegraphics[width=8.5cm]{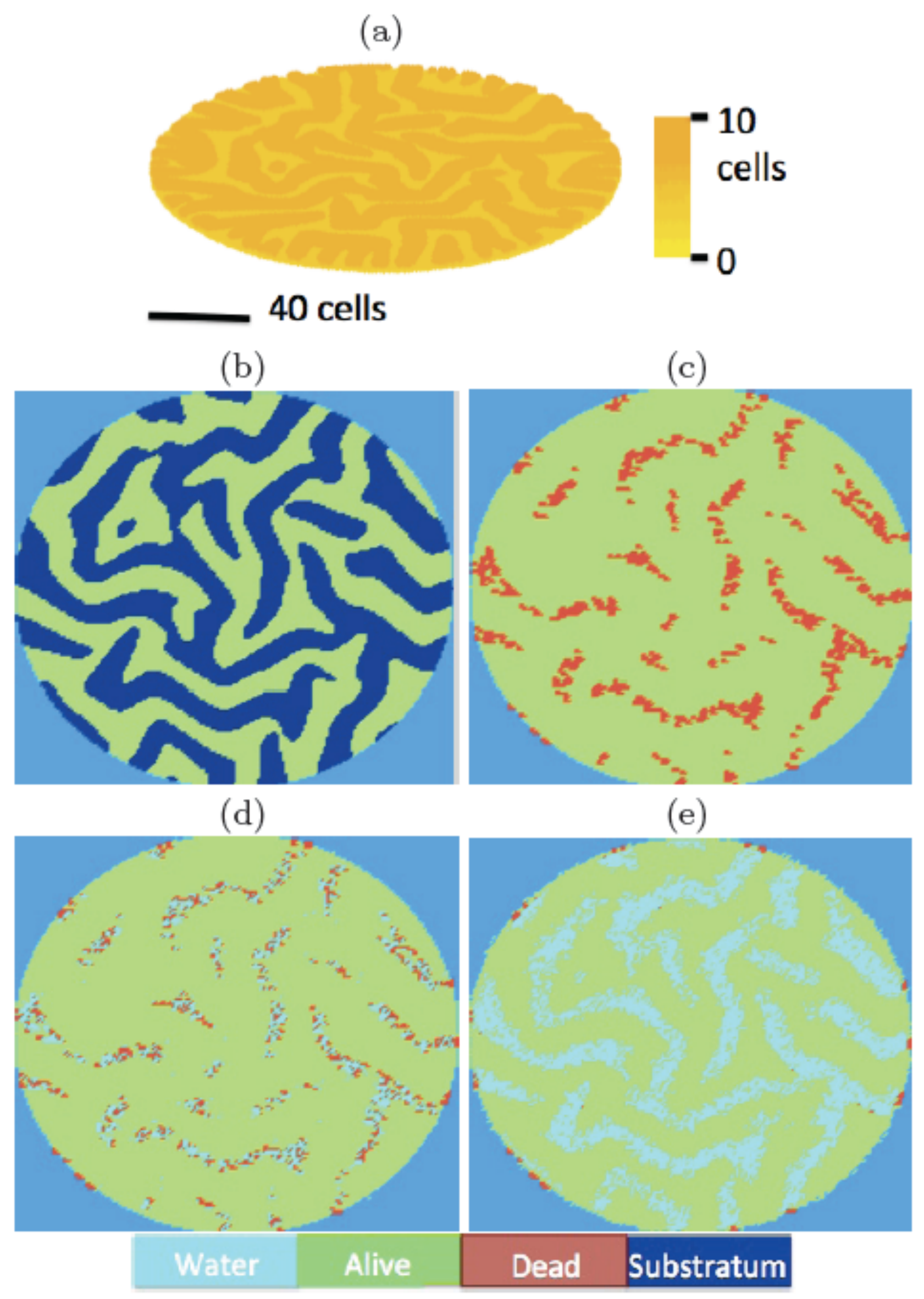} 
\caption{(Color online)
(a) Biofilm-substratum system deformed following out-of-plane displacements
$\xi$ predicted by Eqs. (\ref{plategrowth1bis})-(\ref{plategrowth2bis}) for residual stresses expanding as a radial compression front. 
(b) Bottom slice of (a) displaying the arrangement of biofilm and substratum.
(c) Intemediate slice of (a) showing the allocation of dead cells.
(d) and (e) Accumulation of absorbed water in slice (c) triggered by the 
presence of weakened dead areas after $1$ and $40$ steps of the water absorption process, respectively.
}
\label{fig9}
\end{figure}

In Figure \ref{fig9}, we have distributed dead cells above the substratum, in the wrinkled regions. The value of $\Pi$ is chosen small enough to avoid water proliferation inside the biofilm, about $E/20$. These simple probability laws produce compression residual stresses that vary with the biofilm height, as shown in Figure \ref{fig10}, where $\Pi=E/3$. Values in the range $\Pi \sim E$ are suggested in Ref. \cite{osmotic}.  

{A time scale for water absorption processes may be inferred from the analysis in Ref. \cite{seminara}, where a thin film description of biofilm spread due to swelling and growth is proposed.} For radial biofilms and small departures from osmotic balance, a similarity approximation for the height $h$ and the radius $R$ of the biofilm yields a time scale $1/q$ for the spread process, where $q$ is the rate of biomass production.}

{We cannot directly couple the model in Ref. \cite{seminara} to a cellular automata description of cell division, death and differentiation to describe biofilm spread due to swelling. The main reason is that it already includes a source representing growth. An alternative model of fluid transport in the biofilm would be necessary for a better description of the water absorption process and biofilm spread in our framework.} 

{Once water channels have been carved in the biofilm, we might couple our cellular
automata description to the hydrodynamics of the flow as done for biofilms in porous 
media \cite{kapelloshierarchical}. This assumes that the deformation of the biomass and the extracellular fluid flow can be computed in a sequential manner, which happens when the biomass behaves as an elastic material or the flow induced deformation of the biomass matrix is negligible \cite{kapellosreview}. Using order-of-magnitude analysis,  Ref. \cite{kapellosmultiscale} establishes domains of validity of different models for liquid transport in a fluid-solid system depending on the volume fraction and viscosity of the fluid, the volume fraction, elastic constants and density of the solid, the hydraulic permeability of the system, the characteristic times for displacement of the solid, and the characteristic macroscopic length of the system.}

\subsection{Growth tensor and elastic constants}
\label{sec:growthtensor}

The components of the tensor $g$ entering Eqs.
(\ref{plategrowth1adim})-(\ref{plategrowth2adim}) are the growth 
rates, that is, the rate of volume supply per unit volume. 
New tiles may add to our biofilm due to cell division,  EPS creation or water
adsorption. Wherever they are created, they shift another tile in the direction of
minimal mechanical resistance. The growth tensor at each grid location 
$(ix,iy,iz)$ is computed by keeping track of all the new tiles inserted and the 
direction in which their predecessors where shifted. We define a vector 
${\bf w}=(w_1(ix,iy,iz),w_2(ix,iy,iz),w_3(ix,iy,iz)) a.$ $w_1(ix,iy,iz)$ is determined 
by cumulatively adding $\pm 1$ for each tile shifted in the $x$ direction in the 
positive or negative sense, respectively. $w_2$ and $w_3$ are evaluated in a 
similar way, along the $y$ and $z$ directions. The final vector is normalized
to have norm $a$.
Then, we compute $\nabla {\bf w}$, where the derivatives are 
approximated by finite differences that use the known grid values. To estimate 
$g(ix,iy)$ we consider all the contributions from $\nabla {\bf w}(ix,iy,iz)$ for varying 
$iz$.

\begin{figure}[!ht]
\centering
\includegraphics[width=8.5cm]{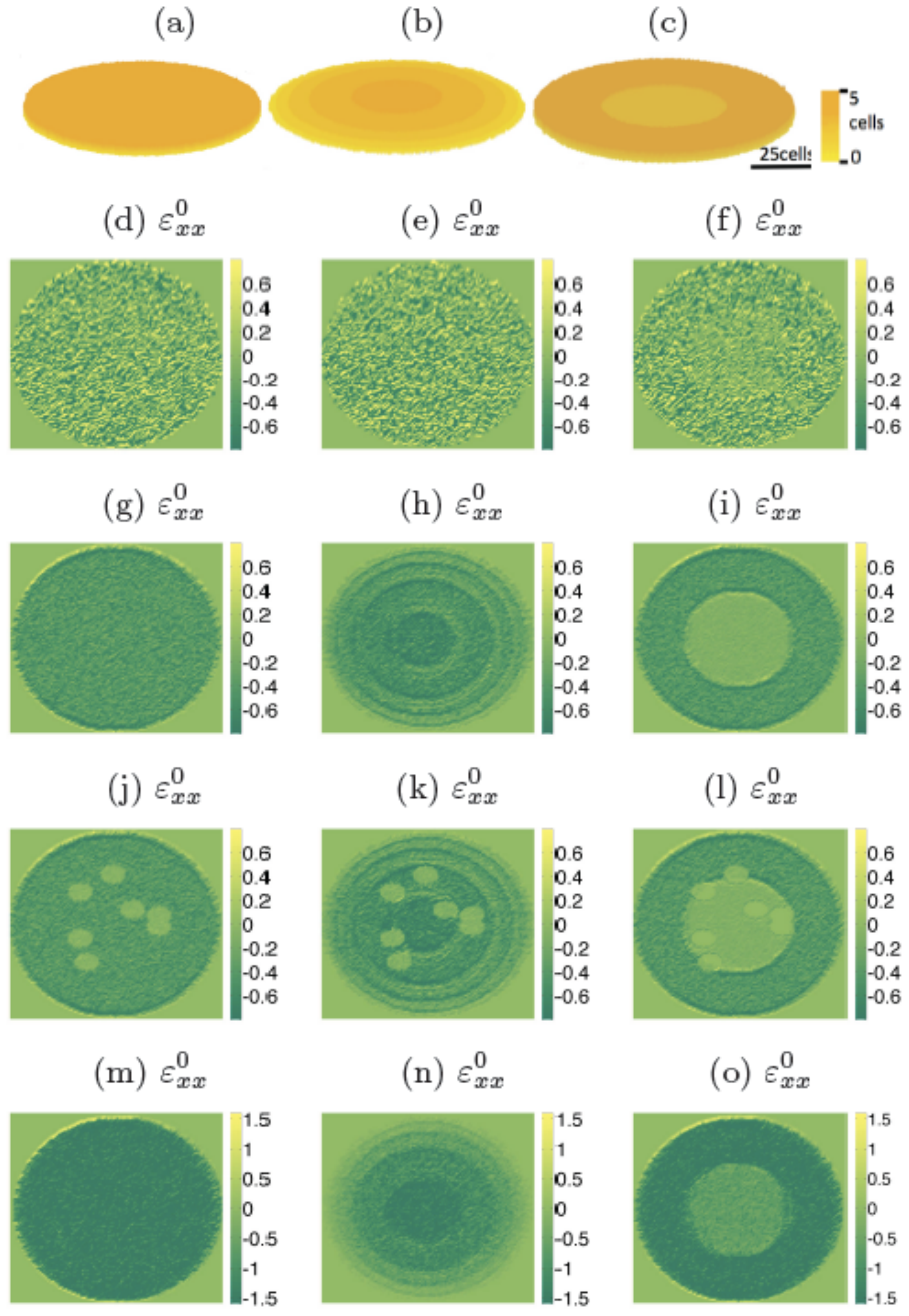}
\caption{(Color online)
Residual stresses $\varepsilon^0_{xx}$ for the biofilm seeds depicted in
(a), (b) and (c). 
(d), (e) and (f) represent their values after one step of the stochastic division 
and spread process, with $C=2 K_n$ and ${k_n a^2 \over D_n K_n}= 0.01$.  
Spatial variations related to thickness are identifiable in (g), (h), (i), obtained averaging a few trials of the stochastic procedure. These fields approach their average values 
(about $-0.1$). A single trial already gives the average value used in rough approximations by constants.
(j), (k), and (l) show the effect of including dead spots in the biofilm seed. Localized
depressions are observed. 
(m), (n), (o) reproduce the residual stresses generated by the water absorption scheme
when $\Pi=E/3$. The average values are about $-0.2$.
$\varepsilon^0_{yy}$ resembles $\varepsilon^0_{xx}$ in all cases. $\varepsilon^0_{xy}$
is about $3$-$2$ times smaller.
 }
\label{fig10}
\end{figure}

The resulting tensor usually presents abrupt oscillations due to stochasticity,
constrained motion along a restricted set of directions in a cubic grid and
varying biofilm thickness. Such oscillations in the grid scale may destabilize 
numerical solutions of Eq. (\ref{plategrowth1adim}). 
Averaging and smoothing these tensors to avoid numerical instabilities, we 
obtain residual stresses that serve as a basis for the numerical tests, 
see Figure \ref{fig10}. 
Averaging rapidly oscillating coefficients and sources is a standard practice to 
study  macroscopic deformations of materials.

The reference Young modulus of the biofilm is estimated from experimental 
measurements \cite{asally}. The knowledge of its microstructure gained from our 
simulations allows to introduce spatial modulations. At each location, we may 
regulate the reference value multiplying it by the sum of weights describing
the status of its neighboring tiles, as we did in section \ref{sec:water}.
A possible qualitative choice for those weights 
are zero for  air, a small fraction of one for dead cells, a larger fraction for 
normal cells and  one for cells producing EPS. The biofilm becomes softer where it 
swells easily. In this way, we obtain spatially dependent elastic moduli that allow to 
investigate variations in the macroscopic deformation of the biofilm in response to 
changes in its microstructure.

\section{Coupling of cellular and mechanical processes}
\label{sec:coupling}

The interaction between the previous descriptions of cellular and
mechanical processes is visualized in Figure \ref{fig4}. The scheme below
details a practical implementation.

\begin{enumerate}

\item{\it Initialization.} A matrix  is created to indicate
the status of the tiles in the computational grid of step $a$. At first, we 
usually set  $S(ix,iy,iz)=2$ in the regions occupied by a biofilm seed, 
assuming that it is formed by undifferentiated cells stuck together. As tiles 
$(ix,iy,iz)$ are filled with differentiated cells, EPS or water in the next steps, 
$S$ will take different positive integer values. Sections occupied by   
air or substratum correspond to zero or negative values.  
This grid is used to keep track of cellular processes and also to discretize
continuum equations. As the biofilm enlarges, we might resort to a coarser 
grid for the latter purpose.
Initially, the concentration of nutrients is set equal to a constant $C$ in the 
substratum and zero elsewhere. The concentrations of waste, ComX, 
surfactin and EPS are set equal to zero everywhere. 

\item {\it Cellular processes.}


{\it (a) Concentration update.} The concentrations of nutrients,
waste and ComX are evaluated solving Eqs. (\ref{cnb})-(\ref{cnl}), 
(\ref{cwb}) and (\ref{ccx}) with the specified boundary and initial 
conditions.  Whenever surfactin or EPS producers are present, 
Eqs. (\ref{cs}) and (\ref{ce}) are solved too. 
All the equations are nondimensionalized as indicated in expressions
(\ref{dimensionlessreaction})-(\ref{dimensionlesstransport}).
The solutions of the time dependent diffusion problems for these 
chemicals relax to their stationary states in a short time, compared 
to the typical time scales for cellular processes. Explicit finite 
difference schemes provide a low cost approximation to the stationary 
concentrations after a number of steps. Those approximations 
are stored and used in the next stages to evaluate behavioral 
probabilities.

{\it (b) Probabilities for cell activities.} 
They are only computed at tiles $(ix,iy,iz)$ occupied by cells.  
Cells that are not already dead are killed with probability $P_w$
whenever the concentration of waste surpasses a threshold 
$\gamma_w$. We evaluate expression (\ref{pdead}) and generate a 
random number  $r\in [0,1]$. If ${\rm max}(r,\gamma_w)<P_w$, we kill 
the cell.  Alive cells deactivate with probability $1-P_d$, $P_d$ given 
by formula (\ref{pdiv}), as long as the nutrient concentration is low enough. 
Active cells become surfactin producers with probability 
$P_s$ defined in formula (\ref{ps}) if the concentration of ComX surpasses
a threshold. Active cells not releasing surfactin become EPS 
producers with probability $P_e$ given by formula (\ref{peps}) when a minimum 
surfactin level is reached and the nutrient concentration is depleted.

{\it (c) Cellular division and spread.} Active cells not secreting 
surfactin divide with probability $P_d$. The newborn cells are placed
in neighboring tiles in the direction of minimal mechanical resistance. 
In early stages, this may be taken to be the shortest distance to the 
biofilm-air interface.  

{\it (d) EPS phase.} Accumulation of EPS may prompt the creation of 
an EPS phase according to Eq. (\ref{pm}).  

{\it (e) Growth tensor and elastic parameters.} 
As EPS production increases, the biofilm acquires consistency. 
We have used in our simulations average measured values of the Young 
modulus. Information on the spatial distribution of different types
of cells allows to spatially modulate that value. Computing the out-of-plane 
deformations in the next stage requires  the previous derivation of a growth 
tensor from the biomass growth process. This is done as indicated in 
subsection \ref{sec:growthtensor}.


\item{\it Mechanical processes.}


{\it (a) Water absorption and water phase.} EPS production changes the
osmotic pressure and prompts water migration from the substratum
towards the biofilm. Water absorption by the biomass may be accounted 
for by  means of formula (\ref{pwater}). Dead zones near the surface will easily fill
with water, which may expand along the wrinkles   
between the biomass and the substratum. A water phase is created.   
{As the volume fraction of water increases and water channels develop,
 a description of liquid flow in the system might be necessary.}

{\it (b) Growth tensor and elastic parameters.}  
Information on the spatial distribution of absorbed water 
allows to spatially modulate the average Young modulus, decreasing 
it in swollen regions. The growth tensor is updated to reflect the 
volume supply due to water absorption.
This is done as indicated in subsection \ref{sec:growthtensor}.

{\it (c) Out-of-plane displacements.} Vertical displacements due to internal stresses 
are estimated solving Eqs. (\ref{plategrowth1bis})-(\ref{plategrowth2bis}). 
The equations are nondimensionalized following Eqs. (\ref{plategrowth1adim})-(\ref{plategrowth2adim}). Basic explicit difference schemes provide low cost 
approximations. Alternatively, faster spectral methods may be used \cite{huang,ni}.
As the biofilm becomes more heterogenous we might need to solve a three 
dimensional elasticity problem or include lower order terms in the Von-K\'arm\'an
equations.

{\it (d) Biofilm deformation.} Tiles $(ix,iy,iz)$ at the interface biofilm-substratum are 
shifted vertically a number of tiles equal to the integer part of $\xi(ix,iy)/a$, pushing 
their neighbors in the three dimensional biofilm. This is done shifting the status of 
such tiles in the matrix $S(ix,iy,iz)$. Relative status of neighbors in the vertical direction 
is preserved. 

\item{\it Iteration.} The biofilm evolution is calculated alternating
the computation of cellular processes up to a time $t_C$ (estimated
from the doubling time), with steps of the water absorption processes 
in a time scale $t_W \sim t_C$ (estimated from biomass production) and 
steps of the deformation processes in a time scale $t_D$ (which requires 
precise measurements of the substratum parameters). 

\end{enumerate}

Summarizing, the mechanical processes change the geometry of
the biofilm. The equations for the different concentrations must be
solved in the new geometry, which affects the cellular processes
altering the probabilities for the different behaviors. The cellular
processes alter the elastic parameters and the pressure in the 
biofilm, contributing residual stresses for its deformation due to 
growth, death and fluid migration.

\section{Simulation results}
\label{sec:results}

Series of simulations were performed to illustrate the behavior of the model and its limitations, 
choosing single mechanisms or combinations of a few of them. 
{The results show wrinkled patterns and cell distributions in qualitative agreement 
with recent experimental observations.}
They provide insight on the influence of  parameters,  cellular activities and  mechanical processes on the biofilm shape and structure.

\begin{figure}[!ht]
\centering
\includegraphics[width=8.5cm]{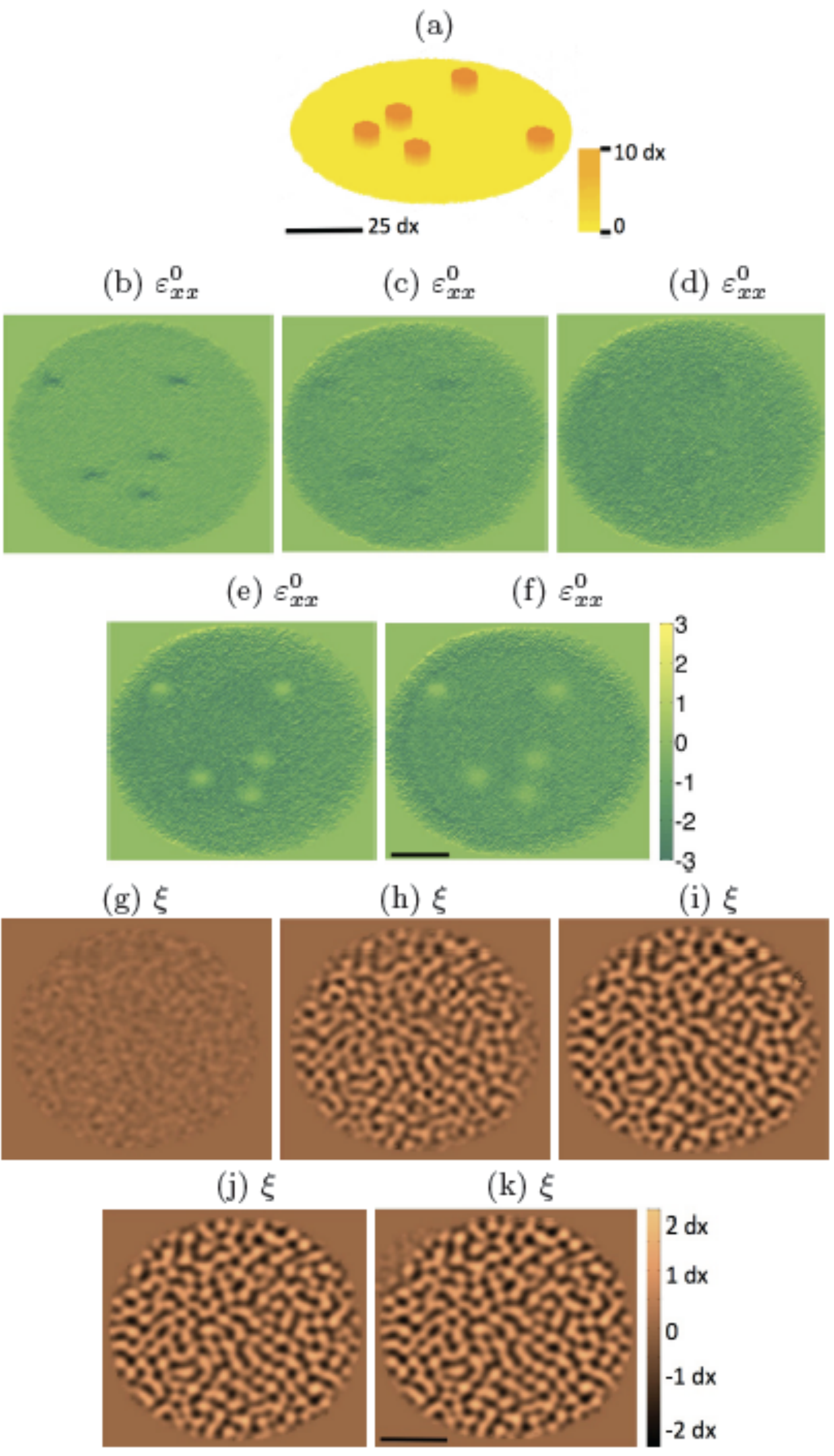}
\caption{(Color online)
(a) Initial biofilm seed.
(b), (c), (d), (e), (f) Residual stresses $\varepsilon^0_{xx}$ averaged after $100$ trials of the stochastic processes at steps 4, 8, 10, 11, 12. The average compression is  $-0.132$, $-0.1425$, $-0.146$, $-0.1437$, $-0.124$. respectively. $\varepsilon^0_{yy}$ has a similar structure and average. $\varepsilon^0_{xy}$ takes smaller values and is neglected.
(g), (h), (i), (j), (k) Vertical displacements. The height mark $dx$ corresponds to one cell. Parameter values are $\gamma=8$, $\nu=0.5$, $E=25\, kPa$, $\nu_v=0.45$ and $\mu_{v}=0$.
 }
\label{fig11}
\end{figure}

Basic growth and spread mechanisms do not produce wrinkled shapes. Even if we activate water absorption processes, the biofilm will spread faster, but no wrinkles appear. 
When elastic deformation mechanisms are incorporated, wrinkles begin to form.
We revisit the simulation described in Figure \ref{fig7} for the same parameter values and a similar initial biofilm seed containing five peaks, see Figure \ref{fig11}(a). During the third step of the growth, spread and differentiation processes, EPS producers appear. At the fourth step, we consider that enough matrix has been produced to regard the biofilm as an elastic film with Young modulus $E$. During the 10-th step, dead cells appear at the bottom of the initial peaks. Dead areas expand during the 11-th step. In the 12-th step, growth in the central region has slowed down due to depleted nutrient levels and increased waste presence.

Figures \ref{fig11}(b)-(f) illustrate the spatial structure of the corresponding  growth tensor. Initial peaks diffuse as the biofilm grows and become depressions when dead cells appear. Once growth in the core is depleted, compression is higher in the border regions. A randomly perturbed initial biofilm deforms according to Eqs. (\ref{plategrowth1adim})-(\ref{plategrowth2adim}) generating small wrinkles that coarsen as time evolves. Figures \ref{fig11}(g)-(k) represent the out-of-plane deformation of the film at selected steps of the growth, spread and differentiation processes. After each step, the average radius of the biofilm increases about one cell. The deformation of the biofilm is then calculated for a time ${\tau \over 24}\, s$.  
As wrinkles develop, the growth processes must be implemented in biofilms with a wavy bottom, contributing additional spatial variations to the growth tensor (higher compression in the valleys, lower compression in the peaks).

 Unless we activate the water absorption mechanism, as in Figure \ref{fig9}, a necrotic region will develop in the biofilm core. Water absorption facilitates nutrient diffusion and waste removal, maintaining the cell normal functions. 

In the computations, we have approximated the residual stresses $\varepsilon^0_{xx}$ and $\varepsilon^0_{yy}$ by their constant average, 
modified by peaks or depressions  that represent increased or decreased compression in certain areas. More accurate smooth approximations can be automatically produced  using denoising strategies borrowed from image processing \cite{marquina}.
As time evolves, singularities may develop due to the presence of `hanging' cells in the border. Such cells are poorly connected to the rest of the biofilm. This artifact is solved by discarding such cells when computing deformations. For the selected parameter values, biofilms are roundish. Therefore, we may alternatively smooth the biofilm border using the averaged support obtained when averaging trials of the stochastic processes to identify spatial variations of the growth tensor. Wavy borders like the ones observed in Figure \ref{fig6}
are usually found for large values of ${k_n a^2 \over D_n K_n}$. This number depends on the nutrient source and the bacterial strain, but tends to be small ( $<< 1$) in practice.

What causes the successive wrinkle branching and the radial branching observed in Figures \ref{fig1}(a) and \ref{fig2}(a)? As we have noticed, EPS production gives the cellular aggregate a certain cohesion. As shown by Figures \ref{fig10} and \ref{fig11}, an expanding biofilm is under compression due to cell division, EPS production and water absorption. The development of wrinkles in Figure \ref{fig11} is limited by size considerations. The radius of the biofilm increases from $60$ to $80$ cells, it thickness is about $10$ cells and $\gamma=8$. Doubling the biofilms maximum radius, and consequently $\gamma$, it appears that a simple round compression front expanding at a certain speed may produce wrinkle branching, structured differently depending on the front speed and the compression magnitude.

\begin{figure}[!ht]
\centering
\includegraphics[width=8.5cm]{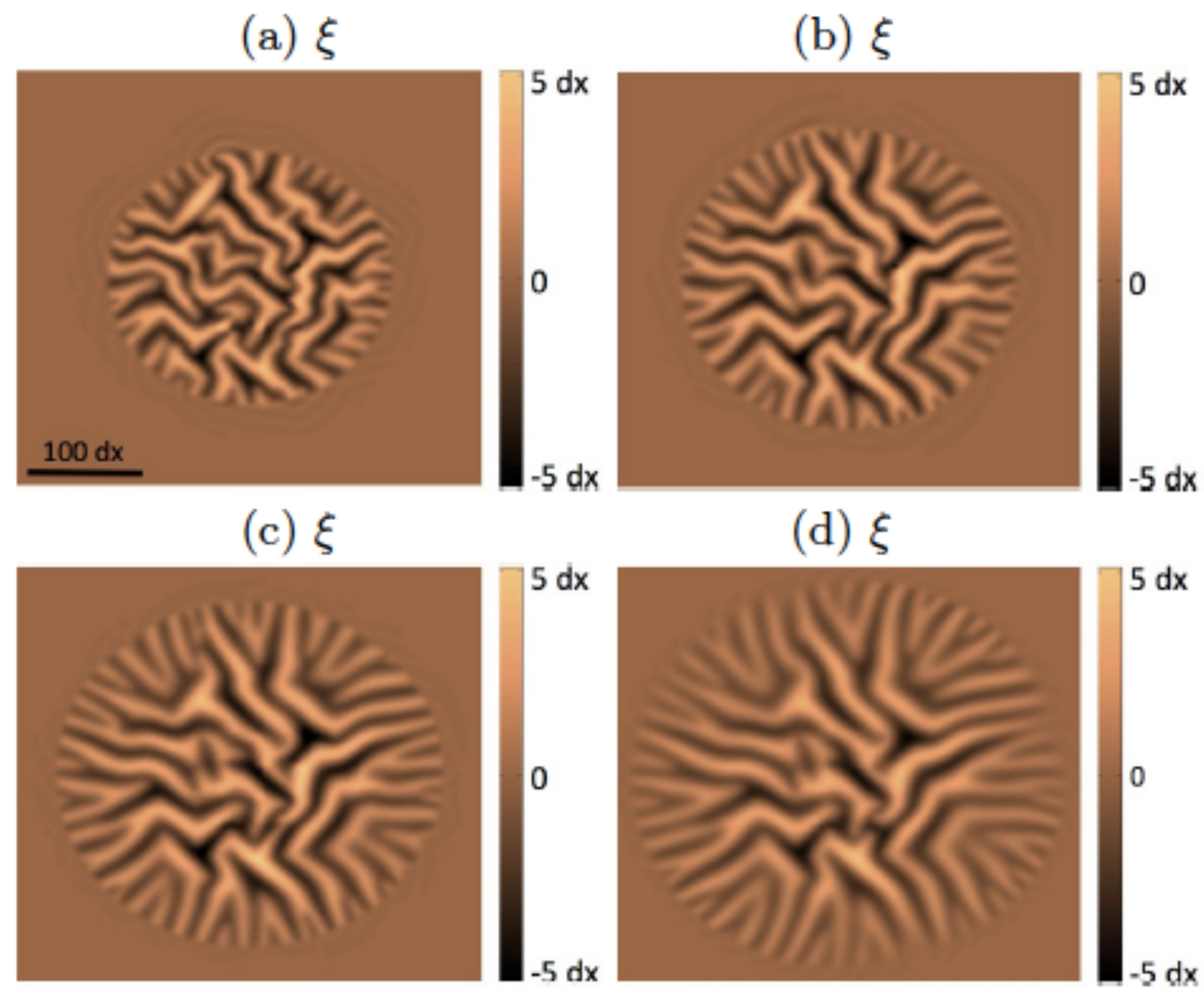}
\caption{(Color online) 
Snapshots of the formation of wrinkles for $\varepsilon_0=0.1$. 
(a), (b), (c), (d) images taken at times $ 280 {T\over \tau}$, $ 560 {T\over \tau}$, $ 840 {T\over \tau}$, $ 1120 {T\over \tau}$. 
An initial random state coarsens to produce a connected wrinkle network that opens up into radial branches as the compressed region spreads. 
The biofilm has Poisson ratio $\nu=0.5$ and Young modulus $E=25\, kPa$. The Poisson ratio and rubbery modulus of the substratum are $\nu_v=0.45$ and $\mu_{v}=0$.
$T$ and $\tau$ are defined in Eq. (\ref{timesdeformation}).
The ratio of the in-plane spatial scale to the out-of-plane spatial scale 
is set to $\gamma=16.$ 
The equations are nondimensionalized so that the dimensionless biofilm thickness becomes $\hat h=1$.  The time step is $dt={0.14 \over \tau} dx^2$. The spatial step is $dx=0.1 \hat h$.}
\label{fig12}
\end{figure}

Motivated by previous observations, the residual stresses $\varepsilon^0({\mathbf x},t)$ are assumed to behave like a radial front $\varepsilon^0(r-vt )$ expanding at a certain speed $v$, $r$ being the distance to the seed center. This allows to lower the computational cost. Figures \ref{fig2}(b) and \ref{fig8} are generated deforming circular biofilm seeds according to Eqs. (\ref{plategrowth1adim})-(\ref{plategrowth2adim}) for such residual stresses. The evolution starts from a configuration with small random vertical displacements. 
Figure \ref{fig8} takes  $\varepsilon^0(r)=-0.1$ constant in the central region. Then, it decreases sharply. The front advances one tile every $14/\tau \; s$. Figure \ref{fig12} shows the time evolution of the wrinkled area. Small wrinkles similar to those in Figure \ref{fig11} form that coarsen as shown in those images. Figure \ref{fig8} is a three dimensional view of
the two dimensional projection depicted in Figure \ref{fig12}(d). 
The biofilm contains scattered dead spots where the residual stresses decrease 
by a certain factor, affecting the way wrinkles nucleate. Once the wrinkled area attains a certain extension, the height of the wrinkles decreases unless we vary
the compression magnitude. If we wish to increase the wrinkle branching rate, while maintaining or enhancing their height as the wrinkled region expands, the magnitude
of the compression has to increase radially. Figure \ref{fig2}(b) was computed raising the speed to one tile every $1.4/\tau \; s$ and increasing the compression magnitude by $r/80$ as the radius of the compressed region grows.
When the compression front expands too slowly, rings may form around the wrinkled area. If it expands too fast, we may see different types of geometric shapes. In between, the network of wrinkles opens up radially as it expands. The outer branches are connected to the inner network and interact with it. 

So far, we have discussed branching of wrinkles. What produces the wrinkled corona in Figure \ref{fig1}(a)? A sudden stop of an advancing compression front may arrange branching wrinkles in a corona type structure temporarily. The whole network changes its structure later.
A uniformly compressed film whose elastic modulus decreases in an outer corona develops stable radial wrinkles, as in Figure \ref{fig1}(b).  The Young modulus $E$ has been decreased by a factor $0.5$ in the outer corona. Radial residual compression of constant magnitude $-0.1$ is applied everywhere, for the same parameter values as in Figures \ref{fig8} and \ref{fig12}. Now, there is no advancing front. The biofilm boundary
remains fixed and the residual stresses are constant everywhere. Radial decrease of the Young modulus might be due to larger water absorption rates in the outer regions, presumably softer, due to the larger presence of normal cells.

\begin{figure}[!ht]
\centering
\includegraphics[width=6cm]{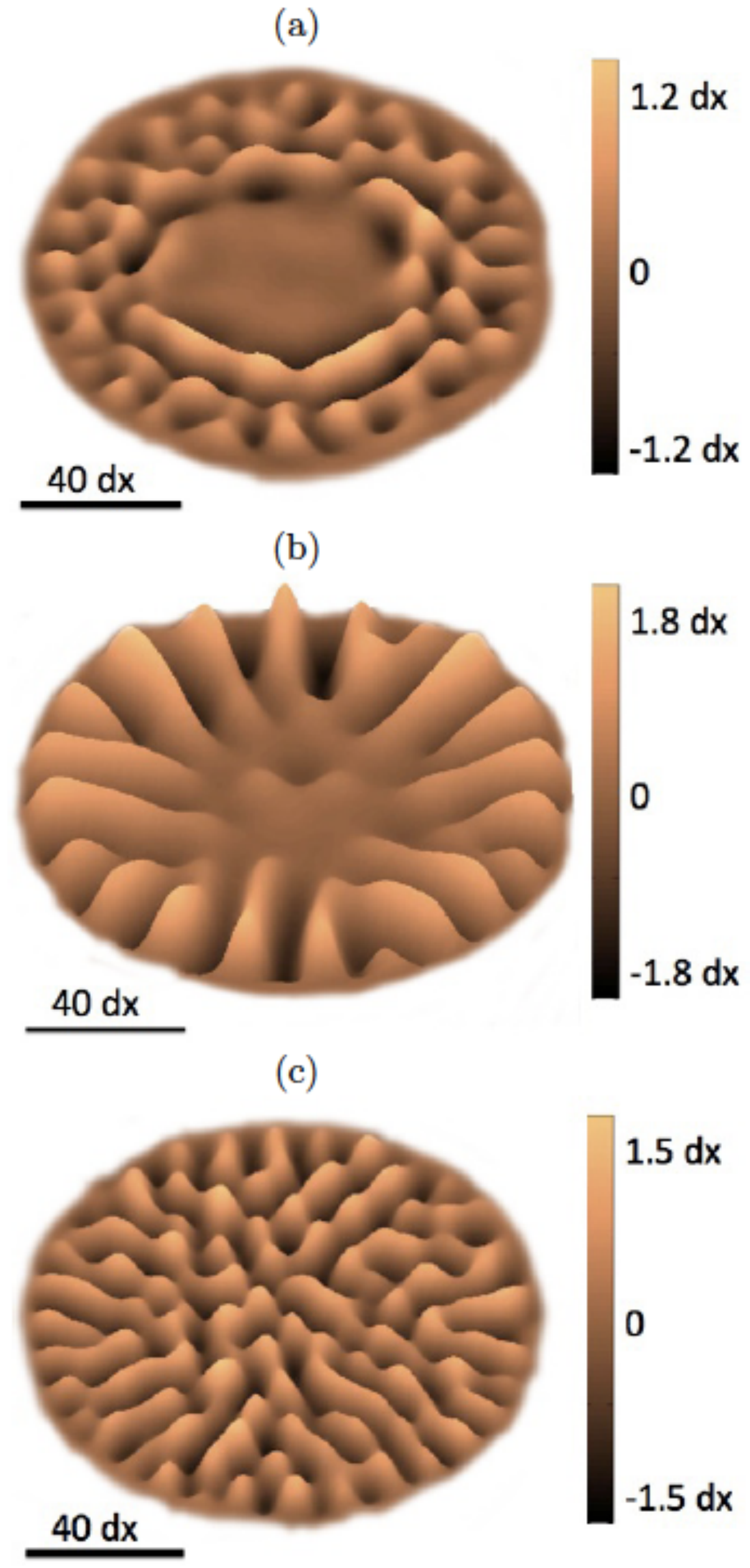}
\caption{(Color online)
Wrinkled structures obtained varying the residual stresses and the elastic moduli for a smaller film with lower Poisson's ratio.
(a) Transition from a harder central region with larger Young modulus to a softer outer corona: radial branches in the corona are separated from the core by rings.
(b) Uniform residual compression in the central region with localized sinks due to dead spots plus radially increasing compression in the outer corona: small wrinkles anchored by the dead spots in the center merge with larger radial wrinkles in the outer corona.
(c) Residual compression increasing slowly with distance to the center: labyrinths formed in the center become radial branches in the outer corona.
Parameter values: Same as previous figures except $\nu=0.3$ and $\gamma=8$.
$dx$ is the spatial step.}
\label{fig13}
\end{figure}

Too sharp a decrease in the elastic modulus, for the selected parameters, combined with smaller values of Poisson's ratio, may result in fast coarsening and disappearance of the central wrinkles, as in Figure \ref{fig13}(a). The compression factor is kept constant in a central core but increases radially in an outer corona. The inner network of wrinkles vanishes with time, but a corona of increasing radial wrinkles is formed.
Both the Young modulus $E$ and the Poisson ratio $\nu$ affect the time scales. Wrinkles coarsen faster in regions with higher Young modulus or lower Poisson ratio.
Central wrinkles, however, are anchored by the presence of a couple of dead spots in Figure \ref{fig13}(b). Decreasing the Poisson ratio enhances height. Lowering the Poisson's ratio to $0.3$, the outer wrinkles not only maintain their heights, but also may increase it for increasing radial compression.  Unlike Figure \ref{fig1}(b) or Figure \ref{fig8}, a radially increasing compression factor $\varepsilon_0$ produces an inner network of wrinkles that opens up forming branches whose  height  can be kept uniform over the biofilm, see Figure \ref{fig13}(c). Notice also that $\gamma$ is decreased by a half.  Radial increase of the compression might be due to larger growth rates in the outer regions caused by increasing availability of nutrients. 

\section{Conclusions}
\label{sec:discussion}

Replicating cell populations create three dimensional  organisms of diverse shapes. Elucidating how those different geometries arise is an intriguing question that has motivated many theories. Small cellular systems, such as bacterial biofilms, seem to develop from the interplay between cellular and mechanical processes.  Studying the interaction of those two mechanisms from an experimental point of view requires concurrent measurement of both processes, posing a major challenge. Developing mathematical and computational frameworks able to incorporate the increasing amount of experimental observations in their governing rules may provide insight on the feedback between microscopic cell behavior and macroscopic continuum processes.
We have proposed a hybrid model to describe the growth dynamics of a small cluster of biofilm on an air-agar  interface.   We are able to transfer information from a stochastic description of cellular processes into a continuum model for the deformation of the film, and use these deformations as a feedback for the cell activities. We have shown that mechanical processes change the geometry of the biofilm. The relevant concentrations must be computed in a new geometry. This fact influences the cellular processes, modifying the probabilities for the cell behaviors. In turn, the cellular processes modulate the elastic parameters and the pressure in the biofilm, and generate residual stresses due to death, growth, and fluid migration. This microscopic feedback determines the subsequent mechanical processes and so on. We have observed  that the properties of the substratum together with the mechanical properties of the biofilm and residual stresses due to death, growth and swelling seem to control wrinkled biofilm shapes.

Our in silico biofilms agree qualitatively with some experimental observations, in the sense that  their shape and the spatial distribution of the different types of differentiated cells is similar. Biomass, air, water and agar distribute through the tiles of a computational grid.
We confer tiles occupied by alive cells the ability to change their status according to probability laws informed by a cascade of concentration fields: nutrients, waste, ComX and surfactin. Local inhomogeneities trigger cell death at the bottom of biofilm peaks. Nutrient depletion deactivates cells at the top. When a threshold level of ComX is reached, surfactin producers appear. When a threshold level of surfactin is achieved, EPS producers proliferate. Once EPS release activates, we consider the biofilm an elastic film, whose elastic parameters are modulated by its microstructure. We also launch a mechanism for water absorption triggered by variations in osmotic pressure caused by EPS production. Weighting at each biofilm location the average values of elastic moduli taken from experiments according to the status of the neighboring tiles, we may produce spatially varying parameters taking into account the microstructure. The biofilm weakens due to the presence dead cells.  It also softens in swollen regions, due to water absorption.
Biofilm expansion is quantified by means of a growth tensor computed from the stochastic growth, death and water absorption processes. In this way, we estimate the residual stresses caused by these mechanisms. The compression residual stresses and the spatially varying moduli used in the numerical tests presented here are motivated by those computations. We study biofilm deformation in response to those stresses by means of a F\"oppl-Von K\'arm\'an approximation, obtaining intricate wrinkled cores that split in radial branches. Water erodes the regions were cells die, expanding along the wrinkles. This improves transport of waste and nutrients, hindering cell death and favoring growth.

Our simulations show that wrinkling seems to be associated with stiffness fluctuations.  Variations in inner residual stresses and elastic constants due to growth, swelling and death seem to govern wrinkle nucleation and branching. 
Compression stresses expanding radially due to swelling and growth produce wrinkled cores that split in radial wrinkles.  The presence of dead regions alters the way the core wrinkles are nucleated. It also favors wrinkle formation and persistence around dead areas as the compression rate is reduced or time increases. Successive wrinkle branching 
may occur depending on the expansion velocity. Radially graded compression stresses may enhance the outer radial wrinkles. Spatially dependent elastic moduli, harder at the center and softer in the outer corona, also enhance the outer radial wrinkles. 

Quantitative comparison with experiments should require careful calibration of several parameters, in particular, regarding substratum properties and  heuristic probability laws. Substratum parameters such as its Poisson ratio, thickness, viscosity and rubbery modulus should be accurately measured since they affect the time scales for the biofilm dynamics. Uptake rates, production rates, saturation constants and diffusivities of the involved chemical compounds should be determined too, since they influence the extend of growth, death and differentiation processes. As the biofilm thickens and its spatial heterogeneity increases, the F\"oppl-Von K\'arm\'an approximation should likely be replaced by a fully three dimensional elasticity model.  {The spread mechanism and the dynamics of water in the system should be revised too.}

\acknowledgments
A. Carpio and D.R. Espeso thank M.P. Brenner for hospitality while visiting Harvard University.
D. R. Espeso and A. Carpio were supported by Comunidad de Madrid and the spanish MICINN through grants No. S2009/DPI-1572, No. FIS2011-28838-C02-02 and No. FIS2010-22438-E. A. Carpio was also supported by a mobility grant of Fundaci\'on Caja Madrid. B. Einarsson was supported by the NILS Mobility project (European economic area-EEA grant) and MICINN grant No. FIS2008-04921-C02-01. Part of the computations of this work were performed in EOLO, the HPC of Climate Change of the International Campus of Excellence of Moncloa, funded by MECD and MICINN. This is a contribution to CEI Moncloa.

\end{document}